\pdfoutput=1
\documentclass[a4paper,twocolumn,10pt]{article}

\usepackage[top=20mm,bottom=20mm,left=17mm,right=17mm]{geometry}
\usepackage[utf8]{inputenc}
\usepackage[T1]{fontenc}
\usepackage{lmodern}
\usepackage{microtype}

\usepackage{amsmath}
\usepackage{amssymb}
\numberwithin{equation}{section}

\usepackage{sectsty}
\allsectionsfont{\raggedright}

\usepackage{graphicx}
\usepackage{booktabs}
\usepackage{adjustbox}
\usepackage[font=small,labelfont=bf]{caption}
\graphicspath{{figures/}}
\numberwithin{figure}{section}
\numberwithin{table}{section}
\renewcommand{\thefigure}{\thesection.\arabic{figure}}
\renewcommand{\thetable}{\thesection.\arabic{table}}

\usepackage{latexml}

\usepackage[hidelinks]{hyperref}

\usepackage[
  backend=biber,
  style=authoryear,
  maxcitenames=2,
  mincitenames=1,
  maxbibnames=10,
  minbibnames=6,
  giveninits=true,
  uniquename=false,
  uniquelist=false,
  sorting=nyt,
  isbn=false, url=false, eprint=false
]{biblatex}

\DeclareSourcemap{
  \maps[datatype=bibtex]{
    \map{\step[fieldset=month, null]}
    \map{
      \step[fieldsource=url, match=\regexp{^https?://(dx\.)?doi\.org/(10\..+)$}, final]
      \step[fieldset=doi, origfieldval]
      \step[fieldsource=doi, match=\regexp{^https?://(dx\.)?doi\.org/(10\..+)$}, replace=\regexp{$2}]
    }
    \map{
      \step[fieldsource=url, match=\regexp{^https?://onlinelibrary\.wiley\.com/doi/(abs/|full/|pdf/)?(10\..+)$}, final]
      \step[fieldset=doi, origfieldval]
      \step[fieldsource=doi, match=\regexp{^https?://onlinelibrary\.wiley\.com/doi/(abs/|full/|pdf/)?(10\..+)$}, replace=\regexp{$2}]
    }
    \map{
      \step[fieldsource=entrykey, match=\regexp{^(Hiraoka2026-ka|Chiba2007-rr)$}, final]
      \step[fieldset=addendum, fieldvalue={in Japanese}]
    }
  }
}
\DeclareFieldFormat{addendum}{\mkbibparens{#1}}
\renewbibmacro*{note+pages}{\printfield{note}\setunit{\bibpagespunct}\printfield{pages}\setunit{\addspace}\printfield{addendum}\newunit}
\renewbibmacro*{addendum+pubstate}{\printfield{pubstate}}
\title{Super-resolved digital terrain models for GEDI geolocation correction:
  agreement with airborne LiDAR DTM-based corrections at the beam level, not the footprint level\thanks{This paper is licensed under CC BY 4.0, except for the images in Figures 4.1, 5.2, and 5.3 that are marked as derived from FABDEM, which are licensed under CC BY-NC-SA 4.0.}}
\author{Natsuki Yuda\thanks{Hitotsubashi University, 2-1 Naka, Kunitachi, Tokyo 186-8601, Japan. E-mail: dm250019@g.hit-u.ac.jp}
  \and Shinichiro Shirota\thanks{Hitotsubashi University, 2-1 Naka, Kunitachi, Tokyo 186-8601, Japan. Corresponding author. E-mail: shinichiro.shirota@r.hit-u.ac.jp}}
\date{}

\newenvironment{labeledlist}{\begin{list}{}{\settowidth{\labelwidth}{(M0):}\setlength{\leftmargin}{\labelwidth}\addtolength{\leftmargin}{\labelsep}\addtolength{\leftmargin}{1em}}}{\end{list}}

\begin{document}

\iflatexml
  \maketitle
  \begin{abstract}
The Global Ecosystem Dynamics Investigation (GEDI) mission measures three-dimensional forest structure with a LiDAR instrument aboard the International Space Station. GEDI observations have geolocation errors of about 10 m, which cause misalignment with other remote sensing data and reduce the accuracy of downstream tasks. Matching GEDI ground elevations against high-resolution digital terrain models (DTMs) from airborne LiDAR is one way to correct these errors, but such DTMs exist in few regions. To widen this method's applicability, we super-resolved a more widely available low-resolution DTM for elevation matching. We ask whether a super-resolved DTM yields the same correction as a 1 m-resolution airborne LiDAR DTM and, if not, why. In steep forests of western Japan, we super-resolved an existing 10 m-resolution DTM to 1 m. Additionally fusing a global 30 m-resolution DTM from a different source substantially improved the accuracy. Using the resulting DTM, we corrected GEDI observation positions first by a single offset per ground track, then point by point. For most tracks, the per-track corrections nearly agreed with those from the airborne LiDAR DTM. Point-by-point corrections, however, did not agree. This was because the error of the generated DTM was close to the elevation differences that point-by-point correction must resolve, and most of this error lay at scales that even a 10 m-resolution DTM can represent. Under similar conditions, a super-resolved DTM may be usable in place of a 1 m-resolution DTM for per-track correction, but improving the accuracy, or using it for point-by-point correction as well, requires reducing the error at these scales rather than reproducing fine relief.

  \end{abstract}

\else
  \twocolumn[{\begin{center}
      {\LARGE\bfseries Super-resolved digital terrain models for GEDI geolocation correction:
        agreement with airborne LiDAR DTM-based corrections at the beam level, not the footprint level\par}
      \vspace{0.8em}
      {\large Natsuki Yuda$^{1}$, Shinichiro Shirota$^{1,*}$\par}
      \vspace{0.3em}
      {\small $^{1}$Hitotsubashi University, 2-1 Naka, Kunitachi, Tokyo 186-8601, Japan\par}
      {\small $^{*}$Corresponding author: shinichiro.shirota@r.hit-u.ac.jp\quad
        (N.~Yuda: dm250019@g.hit-u.ac.jp)\par}
    \end{center}
    \vspace{0.4em}
    \begin{center}
      \begin{minipage}{0.9\textwidth}
        \small
        \noindent\textbf{Abstract}\quad
The Global Ecosystem Dynamics Investigation (GEDI) mission measures three-dimensional forest structure with a LiDAR instrument aboard the International Space Station. GEDI observations have geolocation errors of about 10 m, which cause misalignment with other remote sensing data and reduce the accuracy of downstream tasks. Matching GEDI ground elevations against high-resolution digital terrain models (DTMs) from airborne LiDAR is one way to correct these errors, but such DTMs exist in few regions. To widen this method's applicability, we super-resolved a more widely available low-resolution DTM for elevation matching. We ask whether a super-resolved DTM yields the same correction as a 1 m-resolution airborne LiDAR DTM and, if not, why. In steep forests of western Japan, we super-resolved an existing 10 m-resolution DTM to 1 m. Additionally fusing a global 30 m-resolution DTM from a different source substantially improved the accuracy. Using the resulting DTM, we corrected GEDI observation positions first by a single offset per ground track, then point by point. For most tracks, the per-track corrections nearly agreed with those from the airborne LiDAR DTM. Point-by-point corrections, however, did not agree. This was because the error of the generated DTM was close to the elevation differences that point-by-point correction must resolve, and most of this error lay at scales that even a 10 m-resolution DTM can represent. Under similar conditions, a super-resolved DTM may be usable in place of a 1 m-resolution DTM for per-track correction, but improving the accuracy, or using it for point-by-point correction as well, requires reducing the error at these scales rather than reproducing fine relief.


      \end{minipage}
    \end{center}
    \vspace{1.2em}
  }]

  \begin{NoHyper}\renewcommand{\thefootnote}{}\footnotetext{This paper is licensed under CC BY 4.0, except for the images in Figures 4.1, 5.2, and 5.3 that are marked as derived from FABDEM, which are licensed under CC BY-NC-SA 4.0.}\end{NoHyper}
\fi

\section{Introduction}

The Global Ecosystem Dynamics Investigation (GEDI) is a LiDAR mission aboard the International Space Station designed to measure the three-dimensional structure of forests globally \parencite{Dubayah2020-zd}. Over land between latitudes 51.6$^{\circ}$N and 51.6$^{\circ}$S, GEDI collects observation data, including ground elevation and canopy height, for each laser-illuminated area (footprint) about 25 m in diameter. These data serve as the basis for a wide range of applications, including global canopy height mapping \parencite{Potapov2021-yj,Lang2023-ic} and aboveground biomass estimation \parencite{Dubayah2022-cc}, and contribute to understanding the structure of forest ecosystems and quantifying the carbon cycle.

GEDI footprints have a horizontal positional error known as geolocation error. In Version 2, available since 2021, the horizontal error is approximately 10 m at one standard deviation \parencite{Tang2023-kw}, and even for Version 3, released in 2026, the mean one-standard-deviation horizontal error is reported to be 9.5 m \parencite{Dubayah2026-mu}; this error causes spatial inconsistency between the observation positions reported by GEDI and the actual measurement positions, making it difficult to match GEDI data with field measurements or other remote sensing data \parencite{Tang2023-kw}. Furthermore, it has been reported that geolocation error increases the uncertainty in canopy height estimation and aboveground biomass estimation \parencite{Roy2021-xp,Milenkovic2017-ay}, and that the effect of geolocation error is amplified in areas with topographic relief \parencite{Quiros2021-mu,Cardenas-Martinez2025-qs}. Against this background, correction of geolocation error is required to improve the quality of GEDI data and to enable applications at the scale of individual footprints.

Methods proposed for correcting geolocation error include a waveform matching method based on airborne LiDAR point cloud data \parencite{Hancock2019-wg} and DTM-based methods, which compare high-resolution bare-earth elevation data (a digital terrain model, DTM) with the ground elevation observed by GEDI \parencite{Tang2023-kw,Schleich2023-xz,Li2025-oz}. The latter methods search for the horizontal position at which the difference between the ground elevation observed by GEDI and the elevation read from the high-resolution DTM is minimized, and take that position as the estimate of the footprint position. Hereafter, we refer to the elevation data used for this matching as the reference DTM (also called the reference surface). These DTM-based methods have the advantages of being less affected by seasonal changes in vegetation, because they refer to bare-earth elevation, and of being easier to apply over wider areas, because they do not depend on airborne LiDAR point cloud data. Indeed, \textcite{Li2025-oz} corrected more than 2.7 million footprints in forests using a DTM of approximately 1 m resolution covering 153,670 km\textsuperscript{2} of Japan as the reference surface, and showed that the relative root mean square error (RMSE) of canopy height improved from 35.55\% to 21.07\% in a comparison with the maximum canopy height obtained independently from airborne LiDAR. However, most previous studies of DTM-based methods used high-resolution DTMs, and their applicability is also limited to the regions where such DTMs have been produced. There has also been an attempt using a globally available 30 m-resolution digital elevation model (DEM) \parencite{Tsutsumida2026-bv}, but its evaluation was based on the elevation difference from the DEM used for the correction, and it has not been verified whether correction using coarse-resolution elevation data that contain errors is effective as a correction of geolocation error.

Meanwhile, in the field of remote sensing, super-resolution, which estimates high-resolution elevation data from low-resolution elevation data, has been studied as a means of obtaining high-resolution elevation data, and deep learning-based methods in particular have been actively studied in recent years \parencite{Demiray2021-lc,Zhang2022-uq,Helmy2026-us}. Notably, \textcite{Helmy2026-us} performed super-resolution from a 10 m-resolution DTM to 1 m resolution, a result suggesting that surface generation by super-resolution could overcome the constraint on geolocation correction imposed by the scarcity of high-resolution DTMs. However, a DTM generated by super-resolution contains errors at various scales of the terrain. Because geolocation correction determines the position by matching the elevation of the reference surface against the observed ground elevation, the reference-surface error directly affects the correction results. The question of this study is therefore whether a generated reference surface that contains errors gives the same correction results as a measured 1 m-resolution DTM used as the reference surface and, if not, why not.

In this study, we propose a method that integrates super-resolution of DTMs with GEDI geolocation correction in order to extend geolocation correction to regions where no high-resolution DTM has been produced. The goal is to approximately reproduce, in regions where such data do not exist, the correction results that would be obtained if a 1 m-resolution DTM were available. A 10$\times$ super-resolution model is trained using the 10 m-resolution and 1 m-resolution DTMs published by the Geospatial Information Authority of Japan (GSI). These data represent the bare-earth elevation of the same area, but were produced independently by different measurement methods: the 1 m-resolution DTM from airborne LiDAR and the 10 m-resolution DTM from topographic map contour lines. In addition to the 10 m-resolution DTM, the super-resolution model uses a global 30 m-resolution DTM from which forest and building heights have been removed (FABDEM) as an auxiliary input, and extracts the features of both through intermediate fusion (mid-fusion). Using the generated high-resolution DTM as the reference surface, the per-beam systematic error and the per-footprint individual error are corrected in stages. For the per-footprint search, flow accumulation is adopted with the aim of making the result less susceptible to the instability of the solution inherent in minimizing the cost at a single point, and its results are compared with those of grid search. Footprints that reached the rim of the search disc are excluded as correction failures, and the correction results are compared on the intersection of the footprints remaining in each of the compared results.

As an example of forested areas, the evaluation is carried out in forested areas of western Japan where a 1 m-resolution DTM has been produced. Approximately 330,000 GEDI footprints in 10 test blocks that were not used for training the super-resolution model are the targets of correction. In this study, the 1 m-resolution DTM is regarded as the true terrain. Accordingly, this 1 m-resolution DTM is at once the training target for super-resolution, one of the bases for evaluating the correction, and the basis for defining the reference-surface error. Furthermore, a correction that directly uses the same 1 m-resolution DTM as the reference surface is performed within the same framework and serves as the benchmark for comparing the correction results. Note that this benchmark is itself an estimated corrected position and does not represent the true horizontal position. The effect of the reference surface is examined by the agreement between the corrected positions obtained with the super-resolved reference surface and those obtained with the 1 m-resolution DTM. In addition, both are evaluated by the ground elevation after correction and by the canopy height referenced to an airborne LiDAR-derived canopy height model obtained independently of the reference surface.

The contributions of this study are as follows. First, as the first attempt to apply a DTM generated by super-resolution to geolocation correction, we propose a framework that integrates super-resolution and geolocation correction, and show its potential to extend GEDI geolocation correction beyond the regions where high-resolution DTMs have been produced, to which it has so far been limited. Second, we show that a reference surface obtained by super-resolution from an existing, publicly available low-resolution DTM and an auxiliary DTM gives nearly the same offsets as the 1 m-resolution DTM in the per-beam correction, whereas in the per-footprint correction it does not reproduce the corrected positions obtained with the 1 m-resolution DTM. Third, we show with the error spectrum that most of the super-resolution error consists of components at scales that a 10 m-resolution DTM can represent, and argue that super-resolution useful for geolocation correction requires reducing the error at these scales rather than reproducing fine details.

Super-resolution with only the 10 m-resolution DTM as input left elevation biases that vary from place to place, but fusing the auxiliary DTM reduced the RMSE of the reference surface from 5.32 m to 3.81 m. The offsets of the per-beam correction with this reference surface agreed with those obtained with the 1 m-resolution DTM to within 2.5 m for 56.4\% of the beams and to within 5 m for 84.6\% of the beams, and the agreement was higher for beams with more footprints. In contrast, the per-footprint corrected positions did not agree with those obtained with the 1 m-resolution DTM. When the correction used a reference surface obtained by synthesizing a 10 m-resolution DTM from the 1 m-resolution DTM and training the super-resolution model, the corrected positions obtained with flow accumulation agreed; we therefore consider that this disagreement is due to the error of the generated reference surface. Most of the reference-surface error lay at scales that even a 10 m-resolution DTM can represent.

In the following, Section 2 reviews related work on geolocation correction and super-resolution, and Section 3 describes the data used. Section 4 describes the super-resolution model and the geolocation correction methods adopted in this study. Section 5 presents the super-resolution accuracy and the geolocation correction results, and Section 6 discusses the results obtained in this study.

\section{Related work}

\subsection{GEDI geolocation correction}

GEDI footprints have horizontal positional errors (geolocation errors), which directly affect the accuracy of the data products. These geolocation errors arise from multiple factors, including the pointing accuracy of the lasers, the accuracy of the position and attitude of the International Space Station, and the accuracy of the ranging and time-tag corrections.

Several studies have quantitatively evaluated the magnitude of the geolocation errors. Although the design specification of GEDI gives a horizontal geolocation accuracy within 10 m at one standard deviation \parencite{Dubayah2020-zd}, \textcite{Yang2024-mi} reported that, in northeastern China, the standard deviation of the geolocation deviations of individual footprints was approximately 15 m, exceeding the nominal value. \textcite{Shannon2024-oq} estimated the systematic geolocation error to be 9.62 m (toward the southwest) using a Bayesian model and showed that the probability that the true footprint center lies within 20 m of the reported position is only approximately 45\%. \textcite{Quiros2021-mu} conducted a comprehensive accuracy assessment of 12,031 footprints in southwestern Spain and confirmed that moving the footprints by 10 m in the 270-degree direction reduced the ground elevation error the most. \textcite{Tang2023-kw} reported that 22.7\% of the beams in the GEDI Version 1 data had systematic errors exceeding 25 m, whereas in Version 2, 80.8\% of the beams had systematic errors of less than 10 m.

The influence of geolocation errors on downstream data products has also been evaluated. \textcite{Roy2021-xp} simulated GEDI waveforms from airborne LiDAR in a secondary tropical forest in the Democratic Republic of the Congo and performed Monte Carlo simulations in which the footprint center was moved so as to mimic geolocation errors. The interquartile range of the simulated relative height RH95 varied among footprints from 0.02 m to 17.98 m and was larger in spatially heterogeneous canopies containing small stands, canopy gaps, and forest edges. \textcite{Milenkovic2017-ay} simulated spaceborne LiDAR from airborne LiDAR data and showed that a geolocation error of 14 m increases the variance of aboveground biomass estimates by 17\%. \textcite{Quiros2021-mu} and \textcite{Cardenas-Martinez2025-qs} reported that the influence of geolocation errors is further amplified in areas with steep slopes.

Methods for correcting geolocation errors are broadly classified into two categories according to the type of reference data used. \textbf{Waveform matching-based methods} simulate waveforms that mimic the observation conditions of GEDI from airborne LiDAR point clouds and search for the position at which the similarity to the observed waveform is maximized. The GEDI waveform simulator developed by \textcite{Hancock2019-wg} is the foundation of these methods, and \textcite{Xu2023-qe} used it to detect the systematic geolocation error common to each beam at eight sites in the United States. A method that simulates waveforms with a radiative transfer model \parencite{Mkaouar2025-lh} and a method that uses the agreement with canopy height metrics as the criterion \parencite{Wang2025-bj} have also been proposed. GEDICorrect by \textcite{Corado2026-tk} performs matching at three levels, namely orbit, beam, and footprint, based on airborne LiDAR point clouds. This idea of separating the levels of correction is close to the beam-level and footprint-level corrections; however, its footprint-level correction determines a common offset for approximately 8--10 consecutive footprints, which differs slightly from the footprint-level correction of this study, in which the offset is determined from a single footprint alone. All of these methods require airborne LiDAR point clouds.

\textbf{DTM-based methods} compare a high-resolution DTM with the ground elevation observed by GEDI and search for the position at which the elevation difference is minimized \parencite{Tang2023-kw,Schleich2023-xz,Li2025-oz,Tsutsumida2026-bv}. \textcite{Tang2023-kw} detected and corrected the systematic geolocation error of each beam using a 1 m resolution bare-earth elevation model obtained from the 3D Elevation Program (3DEP) of the United States as the reference surface. They adopted a scheme that weights the elevations within a footprint with a Gaussian distribution and applies the correction as a constant offset per beam. \textcite{Schleich2023-xz} developed the GeoGEDI method using a 1 m resolution digital elevation model (DEM) of France and performed a correction that aggregates neighboring footprints into clusters. The GeoGEDI method is characterized by determining the optimal corrected position using flow accumulation, which achieves a more robust correction than a simple minimum search. \textcite{Li2025-oz} performed a point-level correction, in which more than 2.7 million footprints were corrected footprint by footprint, using a DTM of Japan with approximately 1 m resolution. The point-level correction reduced the RMSE of the ground elevation from 18.42 m to 2.67 m and the relative RMSE of the canopy height from 35.55\% to 21.07\%. \textcite{Tsutsumida2026-bv} proposed SALPA, which replaces the exhaustive search over candidate points on a grid with optimization methods such as L-BFGS-B and a genetic algorithm, and reported that, with 0.5--1 m resolution DEMs derived from airborne LiDAR in Nikko, Japan, and the Landes region of France as the reference surfaces, the elevation difference between the GEDI ground elevation and the reference surface became slightly smaller than with the GeoGEDI method. They also showed that the computation runs with AW3D30, a global 30 m resolution digital surface model (DSM), as the reference surface, pointing toward correction with globally available elevation data alone. However, the evaluation was based on the elevation difference from the DEM used for the correction, and the authors themselves state that the extent to which the reduction in the elevation difference corresponds to the correction of geolocation errors requires further verification. Whether a DEM with 30 m grid spacing can resolve the approximately 10 m geolocation error of GEDI has not yet been verified.

DTM-based methods have the following advantages over waveform matching methods. First, they do not require dense airborne LiDAR point cloud data of vegetation and can be applied over wider areas wherever a DTM is available. Second, because a DTM represents the bare-earth elevation, it is temporally stable and can be used as the reference surface even when its acquisition time differs from that of GEDI. This contrasts with vegetation data, which depend strongly on seasonal changes. On the other hand, major challenges include the following: their applicability is limited to areas where a high-resolution DTM is available; in areas with little relief, the sensitivity of the objective function decreases, making it difficult to identify the optimal position \parencite{Tang2023-kw,Tsutsumida2026-bv}; and, when the optimal position is searched for footprint by footprint, the correction may fit local elevation variations of the reference DTM and exceed the true geolocation error \parencite{Li2025-oz}.

\subsection{Super-resolution}

\subsubsection{Super-resolution in computer vision}

The super-resolution techniques used to enhance the resolution of DTMs and DEMs are based on image super-resolution methods developed in the field of computer vision. Image super-resolution is an inverse problem of estimating a high-resolution image from a low-resolution image, and it has developed rapidly since the introduction of deep learning.

A variety of architectures have been proposed as deep learning models for super-resolution. Since SRCNN by \textcite{Dong2016-vt}, which uses a convolutional neural network, models such as the following have been proposed: SRGAN \parencite{Ledig2017-ft}, which generates perceptually natural images with a generative adversarial network (GAN); EDSR \parencite{Lim2017-nm}, which deepens the residual structure; and RCAN \parencite{Zhang2018-bv}, which nests residual structures within a residual structure and combines them with a channel attention mechanism. By adaptively attending to informative feature channels through the channel attention mechanism, RCAN achieves higher accuracy in restoring high-frequency components than a simple residual network. Modified RCAN \parencite{Helmy2026-us}, on which this study is based, is RCAN modified for the super-resolution of DTMs.

\subsubsection{Super-resolution of DTMs and DEMs}

In applying super-resolution architectures from the computer vision field to DTMs and DEMs, models and loss functions have been designed with consideration of the differences in properties from natural images. Whereas the perceptual quality of textures is emphasized in the super-resolution of natural images, numerical accuracy of elevation values is required for DTMs/DEMs. Moreover, because mean absolute error or mean squared error alone does not sufficiently reflect topographic features such as slopes, ridges, and valley lines, the design of loss functions suited to terrain data has also become an important research topic.

Over the past few years, a variety of deep learning approaches have been proposed for the super-resolution of DTMs/DEMs. Among GAN-based methods, \textcite{Demiray2021-lc} generated 3 ft resolution DEMs from 50 ft resolution DEMs with D-SRGAN, which is based on the SRGAN architecture. Among convolutional neural network-based methods, \textcite{Zhang2022-uq} proposed TfaSR, which combines deformable convolution with a cooperative loss function that takes terrain features into account (elevation loss, slope loss, and watershed-neighborhood loss), and performed super-resolution from 30 m resolution to 10 m resolution. Among residual network-based methods, \textcite{Helmy2026-us} modified RCAN to perform 10$\times$ super-resolution from 10 m resolution to 1 m resolution and showed that the combination of the mean absolute error of elevation and slope loss gives the best results. Among Transformer-based methods, \textcite{Zheng2024-de} proposed TUGAN, which uses a self-similarity Transformer as the generator and a U-Net as the discriminator. In addition, \textcite{Liu2026-jj} performed super-resolution of up to 16$\times$ with GSRMTL, a multi-task learning method that uses high-resolution optical images as guidance and jointly learns DEM super-resolution and semantic segmentation, and \textcite{Yu2025-rc} proposed DSRFlow, which is based on normalizing flows.

In studies comparing GAN-based methods with non-GAN-based methods for super-resolution at large upscaling factors of 4$\times$ or greater, many have reported a tendency for the GAN-based methods to be inferior in elevation accuracy \parencite{Zhang2022-uq,Li2023-fn,Zheng2023-vj,Wang2024-wq}. Although results in which GAN-based methods outperform some non-GAN-based methods have also been reported \parencite{Demiray2021-lc,Liu2026-jj}, no result has shown a clear advantage over non-GAN-based methods in general. Most of these existing studies aimed to evaluate the methods they proposed, and few studies have aimed at the comparison of GAN-based and non-GAN-based methods itself. Nevertheless, the fact that a consistent tendency has been observed across multiple independent studies suggests that GAN-based methods tend to be at a disadvantage in elevation accuracy for super-resolution with large upscaling factors. One reason behind this is that GANs are designed to optimize perceptual quality and prioritize the naturalness of textures over the numerical accuracy of elevation values, so that texture hallucination and checkerboard artifacts tend to occur \parencite{Wu2021-hi}. In the super-resolution of DTMs, the numerical accuracy of elevation matters most, and these side effects accompanying the optimization of perceptual quality directly impair accuracy as topographic distortions. From the above, we consider that, at present, accuracy-oriented architectures are more suitable than GAN-based ones for the super-resolution of DTMs with large upscaling factors.

Finally, we address the issue of training and evaluation data common to the super-resolution of DTMs/DEMs. The great majority of super-resolution studies construct their training and evaluation data by synthesizing low-resolution data through downsampling of high-resolution data \parencite{Helmy2026-us,Demiray2021-lc,Zhang2022-uq,Zheng2023-vj,Li2023-fn,Wang2024-wq,Yu2025-rc}, and studies of super-resolution using real DEM pairs are few \parencite{Wu2021-hi,Zhang2022-uq}. Only recently have studies that handle real data pairs with optical images as auxiliary input begun to appear, such as one that guides a global DEM with aerial photographs to perform bare-earth conversion and resolution enhancement simultaneously \parencite{Cai2025-xt} and one that guides real data pairs of urban DSMs with optical images \parencite{Panangian2024-ec}. The main reasons why synthetic data are widely used are that it is difficult to obtain real data pairs in which the same area is observed with different sensors at different resolutions, that pairs from different data sources can be misaligned horizontally or vertically \parencite{Wu2021-hi,Okolie2022-hs}, and that evaluation is easy with synthetic data because the high-resolution data themselves can be used as the ground truth.

\textcite{Wu2021-hi} pointed out a large gap between real data and synthetic data. They performed super-resolution from the 30 m resolution data of the Shuttle Radar Topography Mission to the 10 m resolution data of WorldDEM and experimentally showed that a model trained on synthetic data does not generalize to real-world super-resolution tasks. They attributed this to the fact that the relationship between a low-resolution DEM and a high-resolution DEM in the real world cannot be reproduced by simple downsampling because of differences in sensor characteristics, data acquisition time, and processing methods. Thus, from a practical standpoint, the need for training and evaluating models with real data pairs is strongly suggested.

\section{Data}

This section describes the data used in this study. This study used the 1 m resolution DTM (DEM1A) and the 10 m resolution DTM (DEM10B) published by the Geospatial Information Authority of Japan (GSI), together with FABDEM, for super-resolution, and performed geolocation correction of the observation data of GEDI, a spaceborne LiDAR mission. In addition, the 50 m resolution JAXA High-Resolution Land Use and Land Cover map was used for land use and land cover classification in Japan, and the 0.5 m resolution DCHMs published by Hyogo Prefecture and Kochi Prefecture were used for validating canopy height.

\subsection{GEDI}

\begin{figure*}[t]
\centering
\includegraphics[width=\textwidth]{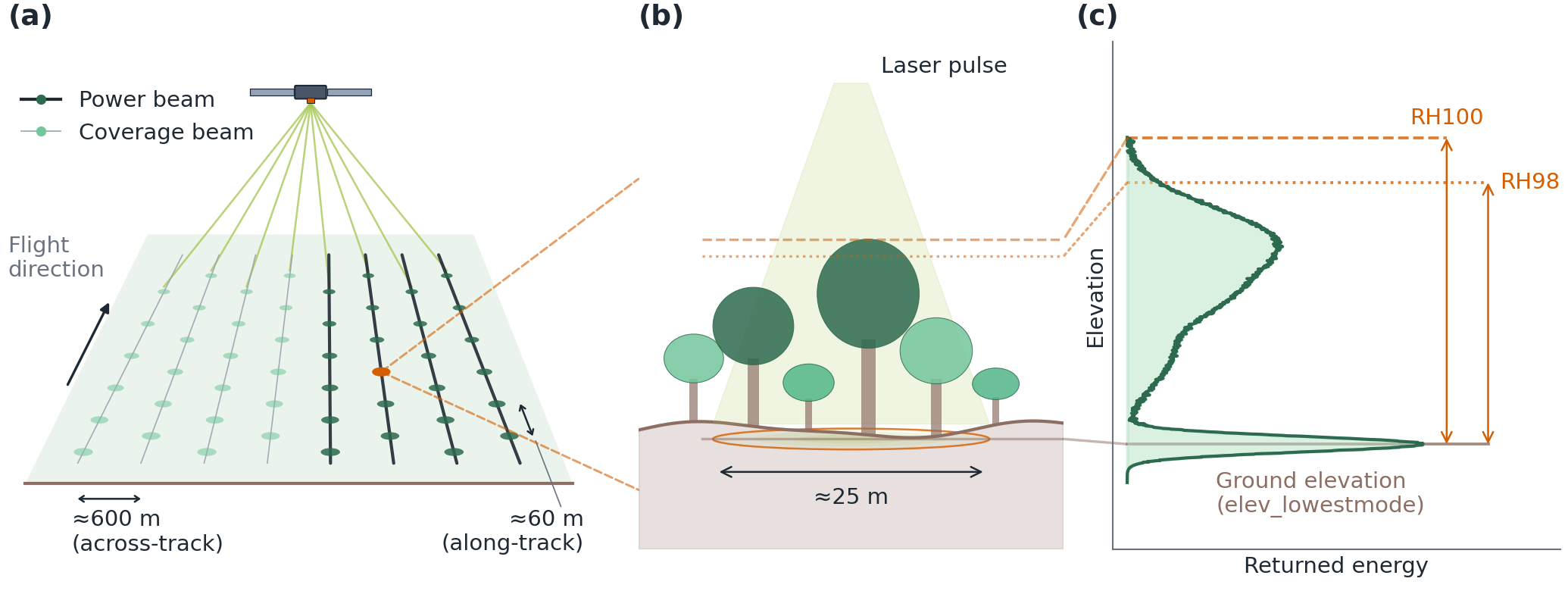}\par\smallskip
\caption{Schematic of GEDI observations. (a) The eight laser beams directed at the ground surface and the tracks they form. Thick lines are power beams, which originate from the high-output lasers, and thin lines are coverage beams. (b) A laser pulse illuminating one footprint and the canopy. (c) The relationship between the received waveform of a footprint and the ground elevation, RH98, and RH100 derived from it. RH98 and RH100 are heights above the ground elevation.}
\label{fig:gedi_observation_schematic_b5faae5c}
\end{figure*}
GEDI is a full-waveform LiDAR mission aboard the International Space Station (ISS), designed to measure the three-dimensional structure of the Earth's forests at high resolution \parencite{Dubayah2020-zd}. GEDI was installed on the ISS in December 2018 and began observations in March 2019. GEDI observes within the latitude range from 51.6 degrees south to 51.6 degrees north. GEDI stopped observing in March 2023 and was stored on the ISS, but it was reinstalled in April 2024 and resumed observations in June of the same year.

The GEDI instrument carries three lasers with a wavelength of 1064 nm, which are optically dithered to produce eight ground tracks. The pulse repetition frequency of the lasers is 242 Hz, and each footprint is approximately 25 m in diameter. Footprints are spaced approximately 60 m apart along track, and the tracks are spaced approximately 600 m apart across track (Fig. 3.1). In GEDI, each of these eight ground tracks is called a beam and is identified by a name from BEAM0000 to BEAM1011. The four beams originating from the two high-output lasers are power beams (BEAM0101, 0110, 1000, 1011), and the four beams originating from the one laser whose output is split in two are coverage beams (BEAM0000, 0001, 0010, 0011) \parencite{Tang2023-kw}. In this study, the term beam refers to the sequence of footprints aligned along one ground track of one orbit.

As a full-waveform LiDAR, GEDI records the entire reflected waveform of the laser pulse, so that the vertical structure of vegetation from the top of the canopy to the ground surface can be obtained in a single measurement. In particular, the \textbf{relative height percentiles} (RH) are metrics obtained by accumulating the energy of the received waveform from the bottom of the waveform and expressing the height at which the cumulative energy reaches a given fraction of the total energy as a height above the ground elevation (the elevation of the lowest mode); RH98 is the height at which the energy reaches 98\%. RH98 corresponds to the elevation near the canopy top minus the ground elevation, and is the measurement on which canopy height estimation by GEDI is based. The data products have been developed and released in stages, and this study used the Version 2 Level 2A product containing the observations acquired from April 2019 through March 2023.

To ensure the reliability of elevation matching by excluding footprints with low waveform quality and those observed outside forests, we followed \textcite{Li2025-oz} and, as the footprints for geolocation correction, first extracted those that satisfied all of the following six conditions: (1) observed by a power beam; (2) beam sensitivity of at least 0.90; (3) valid and not degraded waveform (l2\_quality\_flag = 1 and degrade\_flag = 0); (4) observed at a site classified as forest in v25.04 of the 2020 and 2022 versions of the JAXA High-Resolution Land Use and Land Cover map (HRLULC); (5) RH98 of at least 5 m; and (6) if located at a site classified as deciduous forest, observed during the leaf-on period of day of year 120--270. To exclude footprints of obviously poor quality, we further added condition (7): the absolute difference between the GEDI ground elevation and the DEM10B elevation is less than 60 m. In the 10 test blocks (Section 3.2), 363,277 footprints satisfied conditions (1)--(6), and 0.56\% of them (2,030 footprints) were excluded by condition (7). From the remaining footprints, we excluded those located where the elevation of the reference DTM could not be obtained and those left as the only footprint in their beam, and the resulting 333,902 footprints were used for geolocation correction.

\subsection{DTM}

A digital terrain model (DTM) is an elevation dataset representing the elevation of the bare ground surface, with features such as vegetation and buildings removed. In contrast, a digital surface model (DSM) represents the elevation of the entire surface, including the heights of vegetation and buildings \parencite{Farr2007-pr,Hawker2022-ft}. Many studies on GEDI geolocation correction use DTMs, which do not include the effects of vegetation.

Few DTMs are available at the global scale. The Shuttle Radar Topography Mission (30 m; \cite{Farr2007-pr}) and Copernicus GLO-30 (30 m) are global elevation datasets acquired by interferometric synthetic aperture radar, but both are DSMs and include the heights of vegetation and buildings. As one of the products that create a bare-earth DTM by removing the effects of vegetation and buildings from such DSMs, the 30 m resolution FABDEM, described later, is freely available with global coverage \parencite{Hawker2022-ft}. Among commercial products, WorldDEM Neo DTM by Airbus provides a global DTM at 5 m resolution. At the national or regional level, high-resolution DTMs have been published in some cases. GSI in Japan, the 3D Elevation Program (3DEP) in the United States, and the National Institute of Geographic and Forest Information (Institut national de l'information g\'{e}ographique et foresti\`{e}re) in France each provide DTMs at 1 m resolution.

All three DTMs used in this study are grids aligned with latitude and longitude, and their grid spacings are defined exactly in arcseconds. The grid spacing is 0.04 arcseconds for DEM1A, 0.4 arcseconds for DEM10B, and 1 arcsecond for FABDEM. In our implementation, these grids are handled with integer coordinates in units of 0.04 arcseconds. When a grid spacing defined in arcseconds is converted to distance, it is nearly constant in the north--south direction, whereas in the east--west direction it becomes shorter at higher latitudes. Table 3.1 shows these distances in the study area (30 to 36 degrees north) together with the production method of each DTM.

\begin{table*}[t]
\centering
\small
\caption{Grid spacing and production method of the DTMs used in this study.}
\begin{adjustbox}{max width=\textwidth}
\begin{tabular}{lrrll}
\toprule
DTM & Grid spacing (arcsec) & North--south spacing (m) & East--west spacing (m) & Production method \\
\midrule
DEM1A & 0.04 & 1.23 & 1.00--1.07 & Airborne LiDAR survey \\
DEM10B & 0.4 & 12.3 & 10.0--10.7 & Contour lines of topographic maps \\
FABDEM & 1 & 30.8 & 25.0--26.8 & Removal of forest and building heights from Copernicus GLO-30 \\
\bottomrule
\end{tabular}
\end{adjustbox}
\end{table*}
This study targeted the forested areas of western Japan from 30 to 36 degrees north and from 128 to 135 degrees east, a region where forests extend over steep mountains. The main reasons for choosing this area are that few of its forests are covered with snow in winter and that the proportion of deciduous forest is small. The study area was divided into equal-sized blocks, 10 of which were designated as test blocks for evaluating super-resolution and geolocation correction, and the remaining area was used for training the super-resolution model. Fig. 3.2 shows the arrangement of the patches used for training and testing. The selection of the test blocks and the definition of the patches are described in Section 5.1.1.

\subsubsection{DTMs of the Geospatial Information Authority of Japan (DEM1A, DEM10B)}

\begin{figure}[t]
\centering
\includegraphics[width=\columnwidth]{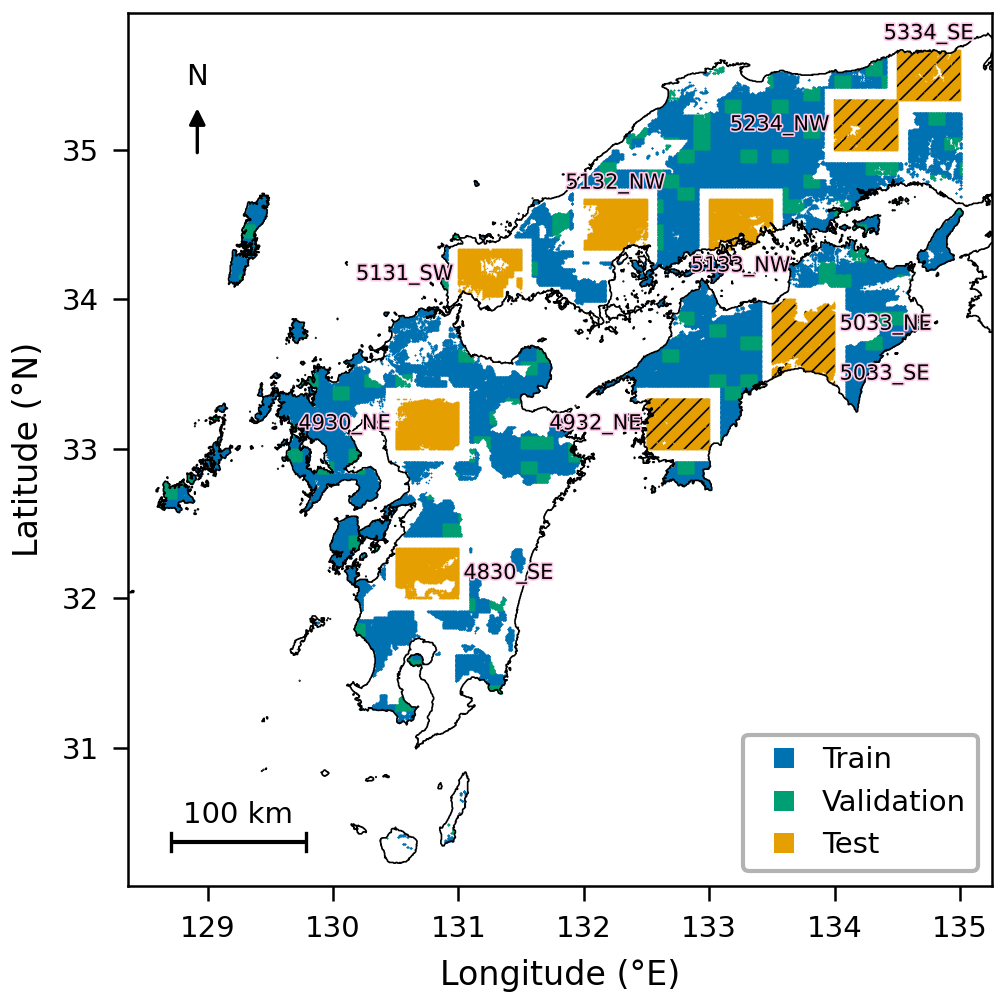}\par\smallskip
\caption{Arrangement of the patches used as Train, Validation, and Test for the super-resolution model. The five test blocks for which DCHMs are available (5334\_SE, 5234\_NW, 5033\_NE, 5033\_SE, 4932\_NE) are hatched. The hatched areas are not the exact areas in which the DCHMs were used. The blank areas around the test blocks are buffers used for neither training nor validation.}
\label{fig:31b25188-e8f2-451f-b38a-1242f880cb1f_0b1eebfd}
\end{figure}
GSI publishes several DTMs with different accuracies and resolutions, two of which were used in this study. DEM1A is a DTM with 1 m resolution based on airborne LiDAR surveys. Its vertical accuracy is within 0.3 m \parencite{Li2025-oz}, and it is the high-accuracy dataset that this study used as the training target of the super-resolution model and for geolocation correction and its evaluation.

DEM10B is a DTM with 10 m resolution, produced from the contour lines of topographic maps. Its vertical accuracy, within 5 m, is lower than that of DEM1A, but DEM10B covers all of Japan and is available even in areas where DEM1A has not been produced. In this study, DEM10B was used as the input (low-resolution data) to the super-resolution model.

\subsubsection{FABDEM}

FABDEM (Forest And Buildings removed Copernicus DEM) is a global 30 m resolution DTM in which the heights of forests and buildings were removed from the Copernicus GLO-30 Digital Elevation Model by random forest regression \parencite{Hawker2022-ft}. FABDEM has been released as the first freely available global DTM at 30 m resolution with forest and building heights removed. In an accuracy assessment over 12 validation areas worldwide, the mean absolute error in forested areas decreased from 5.15 m for Copernicus GLO-30 to 2.88 m, and that in urban areas from 1.61 m to 1.12 m \parencite{Hawker2022-ft}. The horizontal accuracy is inherited from Copernicus GLO-30 and is approximately 4 m.

In this study, FABDEM was used as the auxiliary DTM for the super-resolution model. As far as we could confirm, no elevation data of GSI were used in the source data of FABDEM, in the training data used for removing forest and building heights, or in the explanatory variables, and FABDEM derives from a data source and a production process different from those of DEM10B. Combining the two is therefore expected to reduce bias. Note that the removal of forest heights in FABDEM uses, as an explanatory variable, a canopy height map \parencite{Potapov2021-yj} in which the RH95 of GEDI was estimated wall-to-wall from Landsat imagery.

\subsubsection{Conversion to a common height reference}

The data used in this study have different height references. The elevations of DEM1A and DEM10B are orthometric heights referenced to the mean sea level of Tokyo Bay. The elevations of the DEM1A data used in this study are JGD2024 elevations except in two secondary meshes. A secondary mesh is a cell of the standard regional mesh system of Japan and measures 5 arcminutes in latitude by 7.5 arcminutes in longitude. In the same system, a primary mesh is a cell of 40 arcminutes in latitude by 1 degree in longitude, and a tertiary mesh is a cell of 30 arcseconds in latitude by 45 arcseconds in longitude. DEM10B is not subject to the revision of 31 July 2025, and its elevations are the pre-revision JGD2011 elevations. GEDI elevations are recorded as ellipsoidal heights, and the elevations of FABDEM are, like those of the original Copernicus GLO-30, heights above the EGM2008 geoid \parencite{Pavlis2012-cc}. To unify the height reference, in this study all elevations were converted to JGD2024 elevations before being used for super-resolution and geolocation correction. For DEM10B and for the secondary meshes of DEM1A that had remained in JGD2011 elevations, the amount of the revision was added using the benchmark elevation correction parameters that GSI provides in conjunction with the revision of the official elevations. In some areas for which the benchmark elevation correction parameters are not provided, the triangulation-point elevation correction parameters were used instead. The geoid height of JPGEO2024 (Geoid 2024: Japan and its surroundings; \cite{Hiraoka2026-ka}) was subtracted from the GEDI ellipsoidal heights. For FABDEM, the EGM2008 geoid height was added to the elevations to return them to ellipsoidal heights, and the JPGEO2024 geoid height was then likewise subtracted.

\subsubsection{Terrain roughness}

Many studies have quantified and applied terrain roughness (surface roughness) using DTMs or digital elevation models (DEMs), but there is no single definition. \textcite{Grohmann2011-vl} defined it as ``the variability of a topographic surface at a given scale'' and compared six indices derived from elevation, slope, and curvature. Some indices measure local unevenness, others the relief of the whole area, and others the variability of slope, and which is appropriate depends on the purpose. Moreover, even for the same index, the value changes greatly with the resolution of the source data and the size of the area over which values are aggregated into a single value. \textcite{Grohmann2011-vl} demonstrated this with combinations of five resolutions and 12 window sizes. When comparing roughness, it is therefore necessary to match not only the definition of the index but also the resolution of the data and the size of the area over which the value is aggregated.

In this study, among the indices of \textcite{Grohmann2011-vl}, we used the standard deviation of residual topography ($\mathrm{SD_{restopo}}$) to examine the effect of terrain on super-resolution. Specifically, the terrain roughness of a patch of $500 \times 500$ pixels is defined for the 1 m resolution DTM as follows. Let $z_i$ be the elevations of the valid pixels contained in the region $\Omega$, and $N$ their number. The standard deviation of residual topography is the standard deviation of the residuals obtained by subtracting a smoothed surface from the elevations. Let $W_i$ be the $11 \times 11$ pixel window centered on pixel $i$ and $s_i$ the mean of the valid pixels within it; the smoothed surface is then given by

\begin{equation}
\adjustbox{max width=\linewidth}{$\displaystyle s_i = \frac{1}{|W_i|}\sum_{j \in W_i} z_j$}
\end{equation}
Let the residual be $r_i = z_i - s_i$ and the mean of $r$ within the region be $\mu_r$; $\mathrm{SD_{restopo}}$ is then defined as

\begin{equation}
\adjustbox{max width=\linewidth}{$\displaystyle \mathrm{SD_{restopo}} = \sqrt{\frac{1}{N}\sum_{i \in \Omega}(r_i - \mu_r)^2}$}
\end{equation}
There are two reasons for choosing this index. First, for $\mathrm{SD_{restopo}}$, the residual is zero on a uniformly inclined slope, which is therefore not counted as roughness. This is because this study focuses on the fine unevenness that does not appear in the 10 m resolution DTM, not on the inclination of the slope itself. Second, its values are in meters, so they can be compared side by side with super-resolution errors in the same unit.

\textcite{Grohmann2011-vl} used the same size for the smoothing window and the aggregation area, whereas in this study the smoothing window is fixed at this $11 \times 11$ pixel size and decoupled from the aggregation area (the patch of $500 \times 500$ pixels). The smoothing window was set to $11 \times 11$ pixels because this size is almost the same as the area covered by one pixel of the 10 m resolution DTM given as input. The residual then corresponds to the component that is not contained in the input and must be created by super-resolution.

\subsection{DCHM}

A digital canopy height model (DCHM) is a raster dataset representing the height of vegetation at each grid point, obtained by subtracting a DTM from a DSM. In this study, 0.5 m resolution DCHMs produced by Japanese local governments from airborne LiDAR surveys were used as the data for evaluating whether geolocation correction improves the accuracy of canopy height estimation.

The DCHMs used in this study are those published by Hyogo Prefecture and Kochi Prefecture. The Hyogo Prefecture DCHM was produced from airborne LiDAR survey data acquired by Hyogo Prefecture in fiscal years 2020--2021, and overlaps two test blocks (5334\_SE, 5234\_NW). The same data also include areas surveyed in fiscal years 2012--2013, but these were not used because they do not overlap any test block. The Kochi Prefecture DCHM was produced by Kochi Prefecture from airborne LiDAR survey data acquired by the Forestry Agency of Japan in fiscal year 2018, and overlaps three test blocks (5033\_NE, 5033\_SE, 4932\_NE). Relative to the GEDI observation period (April 2019 through March 2023), the Kochi Prefecture DCHM was surveyed before it and the Hyogo Prefecture DCHM during it, and forest growth or logging that occurred during these time gaps can be a source of error in the comparison between the DCHMs and GEDI RH98.

\section{Methods}

The first half of this section describes the super-resolution method that generates a 1 m resolution DTM from a low-resolution DTM. The second half describes the method for correcting the horizontal geolocation error of GEDI footprints with the generated DTM.

In what follows, the methods are described by the roles that the data play rather than by the names of individual datasets. Super-resolution uses one or more low-resolution DTMs as inputs and a high-resolution DTM as the training target. When there are multiple input DTMs, we call the one with relatively higher resolution and more data the primary input and the others auxiliary DTMs. Of the data described in the Data section, this study uses DEM10B as the primary input, FABDEM as the auxiliary DTM, and DEM1A as the high-resolution DTM. The reasons for incorporating the auxiliary DTM and how it is handled are described in Section 4.1.2.

\subsection{Super-resolution}

As the basis of the super-resolution model, we adopt the Modified RCAN that \textcite{Helmy2026-us} proposed for DTM super-resolution. There are two reasons for this choice. First, that paper deals with the same 10$\times$ super-resolution from 10 m to 1 m resolution as this study, and its configuration is one of the few whose effectiveness has been demonstrated for DTM super-resolution at large upscaling factors. Second, as described in the Related work section, an accuracy-oriented design that restores high-frequency components with residual structures and channel attention suits DTM super-resolution, which requires numerically accurate elevation values.

The configuration of \textcite{Helmy2026-us} has the following three key points. (1) The residual-group structure of RCAN \parencite{Zhang2018-bv} is replaced with a simple body of RCABs (Residual Channel Attention Blocks) connected in series. (2) Reflect padding is used in all convolutions to suppress artifacts at tile boundaries. (3) The 10$\times$ upscaling, which includes a non-integer stage, is split into three stages of 2.5$\times$, 2$\times$, and 2$\times$ and performed by progressive upsampling (Progressive Upsampler), in which each stage consists of bilinear interpolation, a convolution, and a LeakyReLU.

The model of \textcite{Helmy2026-us} takes a single low-resolution DTM as input and is trained and evaluated with low-resolution data synthesized from a high-resolution DTM. This study trains the model with a real DTM pair. Unlike low-resolution data synthesized from a high-resolution DTM, a real low-resolution DTM has errors that originate from its production method. To deal with these errors, this study adopts a model structure that can take multiple DTMs as input. In the following, we denote the primary low-resolution input DTM by $z_{lr}$, the auxiliary DTM resampled to the grid of the primary input by $z_{aux}$, and the high-resolution DTM of the training target by $z$, and their normalized versions by $\bar{z}_{lr}$, $\bar{z}_{aux}$, and $Y$. That is, $z$ and $\hat{z}$ are elevations in meters, and $Y$ and $\hat{Y}$ are elevations in normalized space. Three points are changed or added relative to \textcite{Helmy2026-us}: normalization by the deviation from a base elevation and the Global skip (Section 4.1.1); an encoder for each DTM and mid-fusion to input the auxiliary DTM (Sections 4.1.2 and 4.1.3); and a masked loss and evaluation that exclude missing values (Section 4.1.4).

\subsubsection{Normalization by the base elevation and Global skip}

Unlike the pixel values of natural images, the elevation values of a DEM vary greatly in range from place to place. DEM super-resolution has therefore used methods that normalize the elevation values of each tile to a fixed range \parencite{Zhang2022-ri,Zheng2024-de} and detrending, which takes as the learning target the residual after subtracting a terrain trend obtained by smoothing \parencite{Wang2024-wq}. This study follows the latter idea. The primary input, which consists of $50 \times 50$ pixels, is divided into cells of $10 \times 10$ pixels, and the mean of the primary input within each cell is taken as the base elevation $z_b$. Let $\tilde{z}_b$ be the base elevation expanded to the low-resolution grid by bilinear interpolation, and let $\hat{z}_b$ be $\tilde{z}_b$ further upsampled to the output grid by bilinear interpolation. The inputs and the training target are normalized to deviations from the base elevation: $\bar{z}_{lr}=(z_{lr}-\tilde{z}_b)/k$, $\bar{z}_{aux}=(z_{aux}-\tilde{z}_b)/k$, and $Y=(z-\hat{z}_b)/k$. The constant $k$ is set to 50 (m) in this study. The base elevation $z_b$ of this study is a terrain trend corresponding to the smoothed surface that \textcite{Wang2024-wq} obtained with a Gaussian filter. The Global skip is a connection that enlarges the normalized primary input $\bar{z}_{lr}$ to the output resolution by bilinear interpolation $U$ and adds it to the output of the network $f$, so that the network learns the residual \parencite{Kim2016-im}. What the network learns is the correction added to the bilinearly interpolated primary input; it learns fine relief as well as elevation discrepancies that extend over wide areas. The prediction in normalized space, $\hat{Y}$, and the prediction in meters, $\hat{z}$, obtained by multiplying it by $k$ and adding $\hat{z}_b$, are given by the following equations.

\begin{equation}
\adjustbox{max width=\linewidth}{$\displaystyle \hat{Y}=f(\bar{z}_{lr},\bar{z}_{aux},T)+U(\bar{z}_{lr})$}
\end{equation}
\begin{equation}
\adjustbox{max width=\linewidth}{$\displaystyle \begin{aligned}\hat{z}&=k\,\hat{Y}+\hat{z}_b\\ &=k\,\bigl(f(\bar{z}_{lr},\bar{z}_{aux},T)+U(\bar{z}_{lr})\bigr)+\hat{z}_b\end{aligned}$}
\end{equation}

\subsubsection{Auxiliary DTM}

The purpose of adding the auxiliary DTM to the input is to reduce the bias of the DTM generated by super-resolution. A low-resolution DTM inherently contains systematic errors, and as long as a single low-resolution DTM is the only input, the model can learn these errors only from their correspondence with the training target. If another DTM produced from a data source different from that of the primary input is also given as input, the errors of the two DTMs differ in nature, and part of their biases can be expected to cancel each other out within the model. We therefore require the auxiliary DTM to derive from a data source and a production process different from those of the primary input. However, whether the auxiliary DTM actually contributes to accuracy is not known in advance. An auxiliary DTM usually has a coarser resolution than the primary input and can have errors of several meters in forested areas. Its contribution is something to be verified by experiments that compare models with and without the auxiliary DTM, not an assumption.

The auxiliary DTM is expected to have a grid spacing different from that of the primary input. The auxiliary DTM is therefore resampled to the grid of the primary input by bilinear interpolation before it enters the encoder. In this resampling, missing points among the four neighbors used for bilinear interpolation are excluded, and the interpolated value is computed after renormalizing the weights of the remaining neighbors. Because the reliability of an interpolated value decreases as fewer neighbors are valid, the sum of the interpolation weights of the valid neighbors before renormalization (0 to 1) is attached to the auxiliary DTM as a confidence channel $T$, which tells the encoder how reliable the interpolated value of each pixel is. The encoders consist of stride-1 convolutions and activation functions and output feature maps on the same grid as their inputs, so the outputs of the two encoders are also aligned with the grid of the primary input. In mid-fusion, the outputs of the two encoders are concatenated along the channel dimension pixel by pixel.

\subsubsection{Model architecture}

\begin{figure*}[t]
\centering
\includegraphics[width=\textwidth]{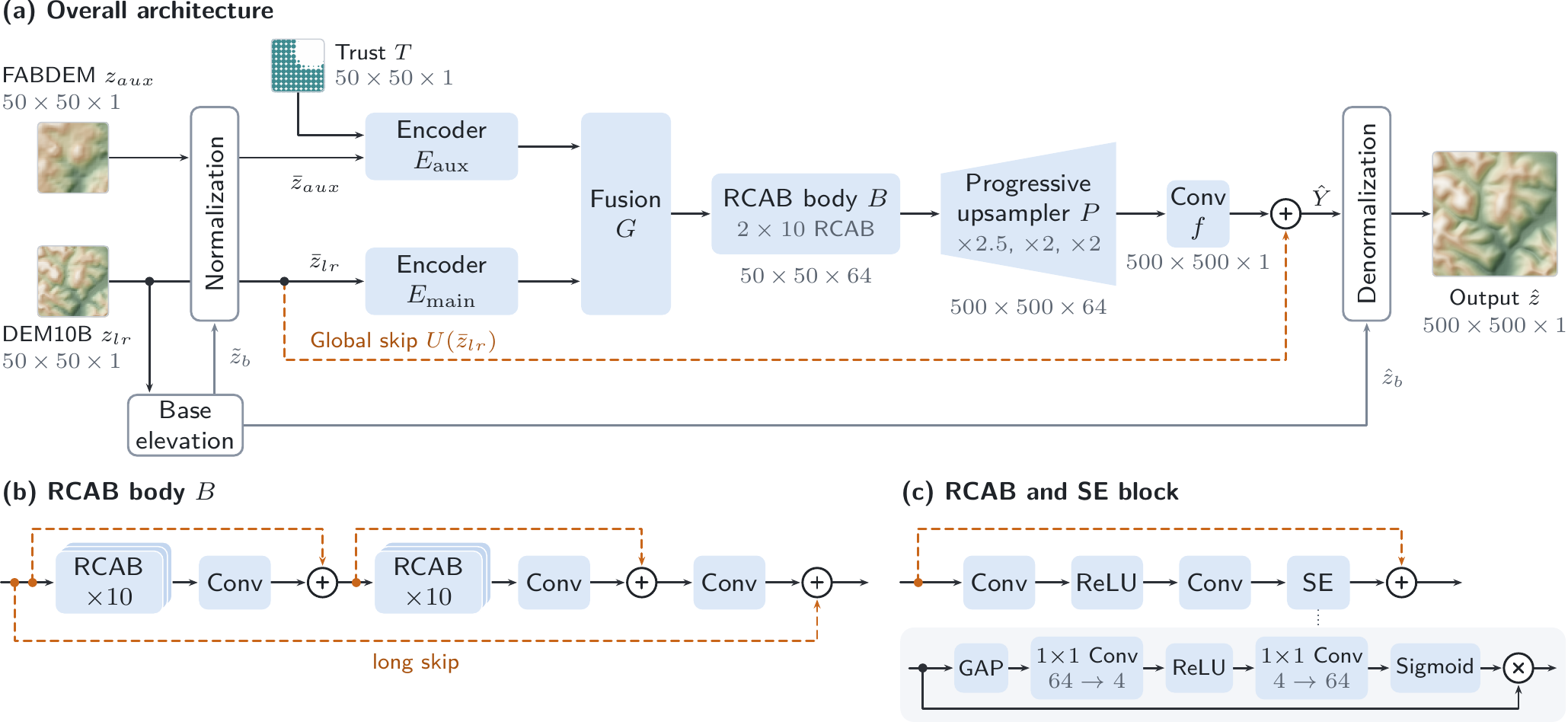}\par\smallskip
\caption{Architecture of the super-resolution model. (a) Overall architecture. (b) RCAB body $B$. (c) Structure of an RCAB and of channel attention (Squeeze-and-Excitation; SE). GAP denotes global average pooling. The details of each layer are given in Appendix A. The FABDEM image in (a) is adapted from FABDEM V1-2 \parencite{Neal2023-fd}, and the output image $\hat{z}$ is derived from it; these two images are licensed under CC BY-NC-SA 4.0.}
\label{fig:61ffd932-da6f-40bf-bd65-7edc8d411cca_71a952fc}
\end{figure*}
The overall architecture of the super-resolution model of this study is shown in Fig. 4.1. The number of channels and the activation function of each layer are given in Appendix A; here we describe the flow of processing. Except for the 1$\times$1 convolutions in the channel attention, the convolutions are 3$\times$3 and use reflect padding. The network $f$ has two input streams, the normalized primary input $\bar{z}_{lr}$ and the normalized auxiliary DTM $\bar{z}_{aux}$ with the confidence channel $T$ attached, both on the grid of the low-resolution DTM. The model of \textcite{Helmy2026-us} has a single input and consists of a single path from the first convolution to the RCAB body, whereas this study provides a dedicated encoder for each input. The primary-input encoder $E_{\mathrm{main}}$ converts $\bar{z}_{lr}$, and the auxiliary-DTM encoder $E_{\mathrm{aux}}$ converts $\bar{z}_{aux}$ and $T$, into feature maps with three convolution layers each. The two feature maps are concatenated along the channel dimension and compressed to the number of channels of the RCAB body by the single convolution layer of the fusion layer $G$ (mid-fusion). When there are multiple auxiliary DTMs, an encoder can be provided for each of them and their feature maps concatenated.

The fused feature map is passed to the RCAB body $B$. $B$ is the part that corresponds to the body of RCAN and consists of 20 RCABs in series. Each RCAB is a residual block that adds to its input the result of a convolution, a ReLU, a convolution, and channel attention applied in this order. Every 10 RCABs are grouped into a residual group with a skip connection, and after the last convolution of $B$, the input of the whole of $B$ is added. This corresponds to the long skip connection of RCAN. We increased the number of RCABs from the 10 of \textcite{Helmy2026-us} to 20 to adjust the capacity to the training data, which are more abundant than in the previous study. The output of $B$ is enlarged to the output grid by the progressive upsampling $P$ in three stages of 2.5$\times$, 2$\times$, and 2$\times$. Each stage consists of bilinear interpolation, a convolution, and a LeakyReLU. Finally, the single convolution layer of the output layer aggregates the features into one channel, which is the output of $f$; adding the Global skip $U(\bar{z}_{lr})$ to it gives the prediction in normalized space $\hat{Y}$ of Eq. (4.1).

\subsubsection{Loss function}

As the loss function, we use the mean of the elevation loss $L_{elev}$ and the slope loss $L_{slope}$, the combination that \textcite{Helmy2026-us} found best. For the training target $Y$ and the prediction $\hat{Y}$, let $M$ be the set of valid pixels, and let $M_x$ and $M_y$ be the sets of positions at which both of the two adjacent pixels needed for the forward differences $(\nabla_x X)_{i,j}=X_{i,j+1}-X_{i,j}$ and $(\nabla_y X)_{i,j}=X_{i+1,j}-X_{i,j}$, respectively, are valid. The losses are defined as follows.

\begin{equation}
\adjustbox{max width=\linewidth}{$\displaystyle L_{elev}=\frac{1}{|M|}\sum_{(i,j)\in M}\left|\hat{Y}_{i,j}-Y_{i,j}\right|$}
\end{equation}
\begin{equation}
\adjustbox{max width=\linewidth}{$\displaystyle \begin{aligned}L_{slope}&=\frac{1}{|M_x|}\sum_{(i,j)\in M_x}\left|(\nabla_x\hat{Y})_{i,j}-(\nabla_x Y)_{i,j}\right|\\ &\quad +\frac{1}{|M_y|}\sum_{(i,j)\in M_y}\left|(\nabla_y\hat{Y})_{i,j}-(\nabla_y Y)_{i,j}\right|\end{aligned}$}
\end{equation}
\begin{equation}
\adjustbox{max width=\linewidth}{$\displaystyle Loss=(L_{elev}+ L_{slope})/2$}
\end{equation}
Because the prediction and the training target are normalized with the same base elevation, each of the losses in Eqs. (4.3) and (4.4) in normalized space is $1/k$ times the value of the same expression evaluated with elevations in meters. Real DTMs contain missing values. The loss and the evaluation metrics are therefore computed only on the set of valid pixels $M$ described above ($M_x$ and $M_y$ for the gradients).

\subsection{Geolocation correction}

In DTM-based geolocation correction, while the position of a footprint is moved horizontally, the elevation at each position is read from a high-resolution DTM and matched against the ground elevation observed by GEDI. In this study, the DTM used for this matching is called the reference DTM or the reference surface. The reference DTMs used in this study include the high-resolution DTM, the 1 m resolution DTMs generated by super-resolution, and the 10 m resolution DTM interpolated bilinearly. All reference DTMs are 1 m resolution raster data, and each pixel has one elevation value representing the area that it covers and the coordinates of the center of that area. The correction searches for the position at which the elevation of the reference DTM best matches the observed ground elevation and moves the footprint there \parencite{Tang2023-kw,Schleich2023-xz,Li2025-oz}. The horizontal vector by which a footprint is moved is called the offset.

The unit for which an offset is estimated differs among studies. \textcite{Tang2023-kw} estimated, for each orbit, one offset common to all footprints of the same beam. \textcite{Schleich2023-xz} estimated an offset for each cluster of about 50 temporally adjacent footprints. \textcite{Li2025-oz} distinguished group-level correction, which gives one offset to each combination of orbit and beam, from point-level correction, which gives an individual offset to each footprint, and applied the two in parallel as mutually independent methods and compared them. In this paper, we call the correction corresponding to the former of \textcite{Li2025-oz} beam-level and that corresponding to the latter footprint-level, and we perform geolocation correction by applying the two stages of correction in series. The flow of the correction is shown in Fig. 4.2.

As for the search algorithms, the correction by grid search and the correction by flow accumulation, described from Section 4.2.2 onward, are combined as follows. First, the beam-level correction is applied by grid search to all footprints, beam by beam. Appendix D shows that using flow accumulation for the beam-level correction hardly changes the corrected positions or the accuracy from those obtained with grid search. Next, the footprint-level correction is performed with the position reached by the beam-level correction ($x_1$) as the origin. At this stage, both grid search and flow accumulation are applied, and their results are also compared. Finally, to see how the correction results differ between methods, four results in total are compared: no correction, correction up to the beam level, and correction up to the footprint level with each of the two methods.

The footprint-level correction starts from the position after the beam-level correction and searches, footprint by footprint, for the geolocation error that remains after the beam-level correction. When the systematic error of a beam is large, a per-footprint search centered on the original coordinates tends to place the true position at or beyond the edge of the search range, which is disadvantageous for the search. Correcting the systematic error of the whole beam first and then searching individually avoids this disadvantage.

\begin{figure*}[t]
\centering
\includegraphics[width=\textwidth]{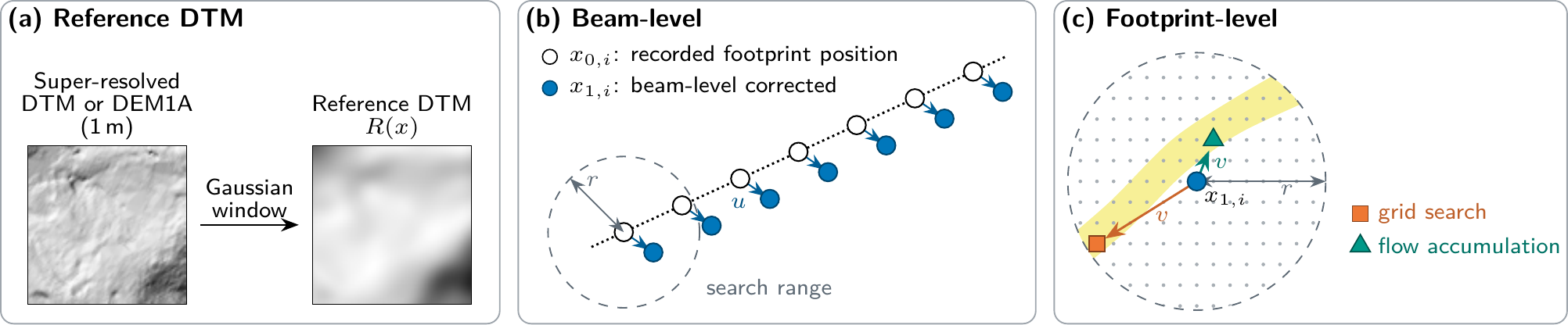}\par\smallskip
\caption{Flow of the geolocation correction. (a) A Gaussian window is applied to the reference DTM to obtain the Gaussian-weighted elevation $R(x)$. (b) Beam-level correction. The dashed circle is the search range of radius $r$. (c) Footprint-level correction. The yellow band represents the region where candidates with elevations close to the observed ground elevation line up along a contour line of the cost. The square and the triangle represent the corrected positions obtained by grid search and by flow accumulation, respectively.}
\label{fig:b0231a96-01e9-479c-9b05-1f6a18b21fdd_28a80393}
\end{figure*}
\begin{figure}[t]
\centering
\includegraphics[width=\columnwidth]{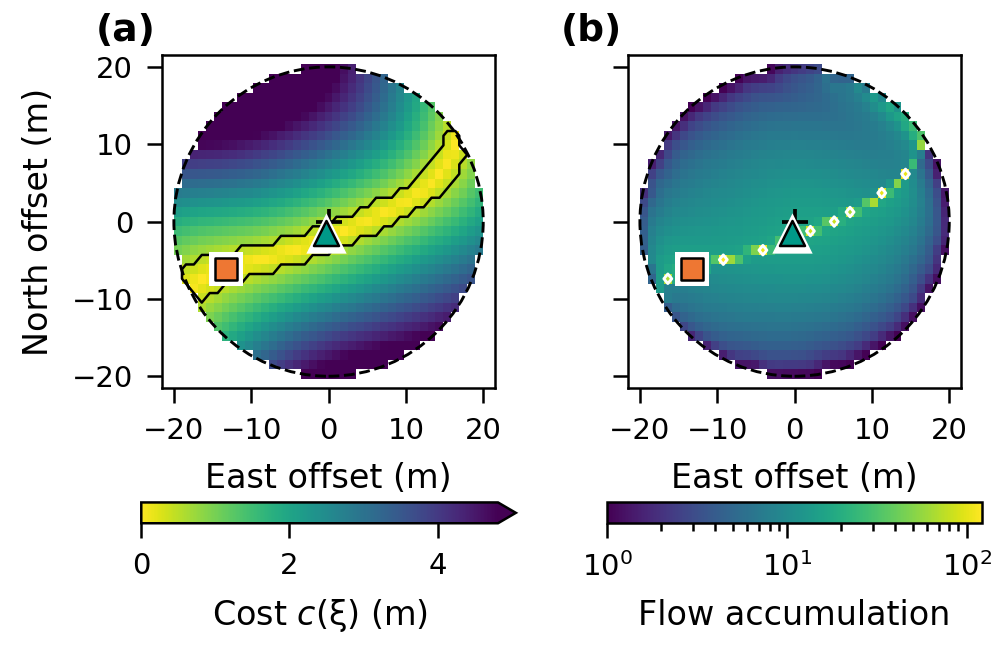}\par\smallskip
\caption{Example of footprint-level correction by grid search and by flow accumulation. The axes are the east--west and north--south distances from the position after the beam-level correction $x_1$ (+), and the dashed line is the search disc of radius 20 m. (a) Visualization of the cost $c(\xi)$. The black line encloses the candidates whose cost is within 0.5 m of the minimum. (b) Flow accumulation computed from the cost map in (a). The white circles show the cells in the top 1\% of flow accumulation. The square is the corrected position obtained by grid search and the triangle is that obtained by flow accumulation; the two are about 14 m apart.}
\label{fig:9c146845-4f12-434d-8269-771616f0d98e_ded59d1c}
\end{figure}

\subsubsection{Gaussian-weighted elevation of the reference DTM and the candidate set}

A GEDI footprint is not a point but a circle about 25 m in diameter, within which the laser energy has a distribution that becomes stronger toward the center. In other words, the terrain observed by GEDI is weighted, and we reflect this weighting in the reference surface used for the correction. The Gaussian-weighted elevation $R(x)$ of the reference DTM at position $x$ is defined as the mean of the pixels of the reference DTM weighted by a two-dimensional Gaussian function centered at $x$ (standard deviation $\sigma=5.5$ m). The reference surface is $R(x)$ viewed as a function of position $x$, and the correction is performed on this surface. This treatment is the same as in \textcite{Tang2023-kw} and \textcite{Li2025-oz}. The weights are determined by the distance between $x$ and the center of each pixel, and $R(x)$ varies continuously with $x$ even within a pixel. The Gaussian-weighted elevation is applied in the search, in the correction results, and in the evaluation alike, regardless of the type of reference DTM.

The candidate set for the corrected position consists of the positions that are offset from the origin in the east--west and north--south directions by integer multiples of the pixel spacing of the reference DTM and whose distance from the origin is within $r$. This study restricts the candidates to discrete points and matches the spacing of the candidates to the resolution of the reference DTM. The origin of the correction is the recorded position $x_0$ at the beam level and the position after the beam-level correction $x_1$ at the footprint level. In this study, $r$ is 20 m at both stages.

Using a disc as the candidate set also differs from previous studies. \textcite{Tang2023-kw} searched a square grid whose axes are the along-track direction of the orbit and the direction perpendicular to it (cross-track). This was based on the assumption that systematic errors originating from attitude and pointing appear along the coordinate system of the orbit. However, the same paper reported that, in the Version 2 data, there was no systematic bias in either the along-track or the cross-track direction and that the offsets were roughly 10 m in magnitude. \textcite{Li2025-oz} searched a square 30 m on a side with east--west and north--south axes in 1 m steps, and \textcite{Schleich2023-xz} also searched a $\pm$50 m square with east--west and north--south axes in 2 m steps. The point-level offsets reported by \textcite{Li2025-oz} showed no discernible tendency in their distribution by direction, and there is no reason to favor a particular direction for footprint-level errors. This study adopted a disc, which does not depend on direction, as the candidate set. This study also identifies correction failures by the Euclidean distance of the displacement, that is, by a circle.

\subsubsection{Correction by grid search}

Geolocation correction by grid search computes, for each candidate in the candidate set, a cost that represents the degree of agreement with the observed ground elevation, selects the candidate with the minimum cost, and takes the vector from the origin to that candidate as the offset. Except for the choice of the candidate set and the classification of correction failures described later, this procedure is exactly the same as the correction of \textcite{Li2025-oz}. With $\xi$ the offset from the origin to a candidate, the cost of one footprint is the absolute difference between the value of the reference surface at the candidate position and the observed ground elevation $z_{obs}$. In the beam-level correction, the origin is the recorded position $x_0$, and the $\xi$ that minimizes the mean of the costs $c_i(\xi)$ of the footprints $i$ belonging to the beam is taken as the offset $u$ common to the beam.

\begin{equation}
\adjustbox{max width=\linewidth}{$\displaystyle c(\xi)=\left|R(x_0+\xi)-z_{obs}\right|$}
\end{equation}
\begin{equation}
\adjustbox{max width=\linewidth}{$\displaystyle \bar{c}(\xi)=\frac{1}{n}\sum_{i=1}^{n}c_i(\xi)$}
\end{equation}
The position after the beam-level correction is $x_1=x_0+u$. When grid search is used for the footprint-level correction, the $\xi$ that minimizes $c(\xi)=\left|R(x_1+\xi)-z_{obs}\right|$ is found with $x_1$ as the origin and taken as the offset $v$ of that footprint, and the position after the footprint-level correction is $x_2=x_1+v$.

\subsubsection{Correction by flow accumulation}

The cost map of a single footprint has multiple candidates with costs close to the minimum, and they also appear at locations far apart from one another. Because grid search selects only the single minimum, a slight error changes which of these candidates is selected (Fig. 4.3(a)). This study also adopts flow accumulation as a more robust method and always compares it with the results of grid search.

Flow accumulation is a way of selecting the corrected position based on the flow accumulation of hydrological analysis, introduced into GEDI geolocation correction by \textcite{Schleich2023-xz}. The cost map $c(\xi)$ is regarded as terrain, the same amount of water is placed on every cell, and the water is routed toward lower cost with the FD8 algorithm (multiple-direction flow to the eight neighbors). Flow collects in regions of low cost, and a region whose cost is consistently low, including its surroundings, receives more flow than a minimum that is a single deep pit. The flow-weighted centroid of the cells in the top 1\% of flow is computed, and the position of the centroid is taken as the corrected position (Fig. 4.3(b)). The choice of the top 1\% is the same as in \textcite{Schleich2023-xz}.

The flow accumulation of this study differs from that of \textcite{Schleich2023-xz} in what it is applied to. That paper applied flow accumulation to an error map averaged over a cluster of about 50 temporally adjacent footprints, whereas this study applies it to the cost map of a single footprint after the beam-level correction. Using the continuous value of the centroid directly as the corrected position is the same as in that paper.

\subsubsection{Identification of correction failures}

In geolocation correction, the wider the search range, the higher the accuracy measured by the agreement between the reference surface and the observed ground elevation. This is because more candidates make it easier to find a location whose value is close to the observed ground elevation. However, this agreement does not mean that the correction is correct. Because it is not possible to verify directly whether the correction of an individual footprint was correct, we infer which corrections are of low quality from the correction results themselves. For example, the larger the displacement, the stronger the suspicion that the cost minimization was driven by an error in the reference DTM or in the observed ground elevation rather than by the true position, and the lower the quality is assumed to be. In particular, for a footprint whose solution reached the rim of the candidate set, the reference DTM or the observed ground elevation may deviate greatly from the true value. We therefore exclude some of the footprints of doubtful quality as correction failures to maintain the quality of the correction results. Previous studies made similar exclusions. \textcite{Tang2023-kw} excluded footprints whose elevation difference from the reference DTM exceeded 15 m as outliers, and \textcite{Li2025-oz} treated the cases in which the optimum fell on the boundary of the search range as failures.

The criterion for how much movement is accepted as a successful correction directly changes the usefulness and the accuracy of the correction results, in that a stricter criterion increases the accuracy of the remaining footprints but reduces the number of usable footprints. In the classification of correction failures in this study, footprints whose solution reached the rim of the search disc are excluded, and only the remaining footprints are kept as evaluated footprints. A solution is said to have reached the rim when the selected candidate, in grid search, or the candidate nearest to the centroid, in flow accumulation, lies on the rim of the candidate set. At the beam level, the unit of exclusion is the beam, so multiple footprints are excluded at once. Because the footprint-level correction is performed after the beam-level correction, footprints excluded at the beam level do not remain among the evaluated footprints of the footprint-level correction either. In addition, because a beam that extends across the test blocks defined in this study is treated as separate beams, some footprints of the same beam may be excluded while others are not. Footprints that have no offset, for reasons such as the reference DTM required for the search being unavailable or there being no valid solution in the disc, are also excluded from the evaluated footprints. When the results of multiple corrections or reference DTMs are compared, the intersection of the footprints that remained as evaluated footprints in each of the compared results is used as the evaluated footprints. This is because, if the evaluated footprints were determined separately for each compared result, the accuracy of the remaining footprints would appear higher for a result that excludes more footprints. For each experiment in the Results section, we state the combination of results whose intersection is used as the evaluated footprints.

\subsubsection{Evaluation}

Geolocation correction is evaluated from two viewpoints, ground elevation and canopy height. The specific metrics are described in the Results section. The evaluation by ground elevation uses the difference between the Gaussian-weighted elevation of the high-resolution DTM at the corrected position and the ground elevation of GEDI. However, this difference has limitations as an indicator of positional correctness. In particular, when the reference DTM used for the correction is itself the high-resolution DTM used for the evaluation, the elevation difference evaluated after the footprint-level correction by grid search is nothing other than the quantity that this correction minimized among the candidates. \textcite{Li2025-oz} reported that, when point-level correction was performed with a 1 m resolution DTM, the ground elevation error became almost zero, with a mean absolute error (MAE) of 0.08 m and a bias of 0.07 m, and referred to this as ``overfitting''. That is, even if the elevation difference after the correction is small, this does not mean that the footprint has moved to its true position. An independent evaluation by canopy height is therefore also needed.

In the evaluation by canopy height, RH98 measured at each GEDI footprint is compared with the maximum of the DCHM within a radius of 12.5 m from the corrected position. RH98 is used rather than RH100, which corresponds to the very top, because it has fewer outliers than RH100 and better represents the canopy height within the footprint \parencite{Hofton2000-cj,Li2025-oz}. The radius is half the footprint diameter of about 25 m, and comparing the maximum within a circle of the size of the footprint with RH98 is the same treatment as in \textcite{Schleich2023-xz} and \textcite{Lang2023-ic}. If the horizontal position has an error, the footprint is placed over a crown different from the actual one, so a reduction in the canopy height difference reflects positional correctness to some extent. Canopy height can also have errors due to differences in acquisition time and measurement errors, but it is data not used in the correction process, and it is used for comparisons between reference DTMs and between correction methods.

As stated at the beginning of Section 4.2, this study compares four results in total: no correction, correction up to the beam level, and the two footprint-level methods. In the evaluation by elevation, the evaluated footprints are nearly the same between correction methods, so the results can be compared intuitively; in the evaluation by canopy height, however, only footprints that have a DCHM value of at least 2 m within a radius of 12.5 m can be evaluated, and the numbers of evaluated footprints do not match exactly between methods either.

\subsection{Relationship between the reference-surface error and the correction}

A reference surface generated by super-resolution has errors that differ from location to location. In this subsection, the high-resolution DTM (DEM1A) is regarded as the true terrain, and the reference-surface error is defined. How the reference-surface error affects the correction is examined separately for the beam level and the footprint level.

\subsubsection{Reference-surface error and the spatial scale relevant to the correction}

The \textbf{reference-surface error} $e(x)$ at position $x$ is defined as the difference between the reference surface $R(x)$ and the Gaussian-weighted elevation $R_{DEM1A}(x)$ of the high-resolution DTM.

\begin{equation}
\adjustbox{max width=\linewidth}{$\displaystyle e(x)=R(x)-R_{DEM1A}(x)$}
\end{equation}
We focus on $e(x)$, weighted by the Gaussian window, rather than on the per-pixel error of the reference DTM in order to discuss the error of super-resolution within the framework of geolocation correction. In the weighted $e(x)$, errors at scales sufficiently smaller than the window width, such as those due to small relief at a scale of a few meters, cancel out. In other words, the per-pixel error of super-resolution consists of errors that appear in $e(x)$ and errors that exist in terms of super-resolution accuracy but are weakened by the averaging over the Gaussian window and have little influence on the matching cost.

The reference-surface error changes the matching cost and can move the corrected position. Taking the origin as $x_0$ and substituting $R(x)=R_{DEM1A}(x)+e(x)$ into Eq. (4.6) gives

\begin{equation}
\adjustbox{max width=\linewidth}{$\displaystyle \begin{aligned}c(\xi)&=\left|R_{DEM1A}(x_0+\xi)+e(x_0+\xi)-z_{obs}\right|\\ &=\left|R_{DEM1A}(x_0+\xi)-\bigl(z_{obs}-e(x_0+\xi)\bigr)\right|\end{aligned}$}
\end{equation}
That is, the reference-surface error enters the cost by changing the elevation to be matched at each candidate by $e(x_0+\xi)$, and the size of this change differs from candidate to candidate. On this basis, the effects of the reference-surface error on the beam-level and footprint-level corrected positions are examined in turn below.

\subsubsection{Effect on the beam-level correction}

In the beam-level correction as well, the cost $c_i(\xi)$ of each footprint contains the reference-surface error, as in Eq. (4.9). However, the beam-level corrected position is the minimum not of the cost of an individual footprint but of the mean $\bar{c}(\xi)$ of the costs of the $n$ footprints in the beam. Because the sequence of the observed ground elevations of many footprints is matched against the terrain of the reference surface at once, we consider that the sensitivity of the offset to the reference-surface errors, which differ from footprint to footprint, decreases as $n$ increases.

If it were possible to estimate how much the reference-surface error moves the offset, the accuracy of the reference DTM required for the beam-level correction would be known. However, such an estimate is difficult for several reasons. First, the reference-surface errors are not independent between footprints but have a spatial structure. Furthermore, because $\bar{c}(\xi)$ is a mean of absolute values, the effect on the offset is not determined by the mean of the errors alone. In addition, the terrain has a complex structure within the search disc, and the reference-surface error is not uniform within the search disc either. Quantifying the effect on the offset while taking these into account is beyond the scope of this study.

In Section 5.4.1, we check directly whether the reference-surface error changes the beam-level corrected position, by examining whether the beam-level offset $u$ estimated with the super-resolved reference DTM agrees with that estimated with the high-resolution DTM. The degree of agreement is also shown according to the number of footprints $n$ in the beam. In Section 5.4.2, we show the mean absolute value and the standard deviation of the reference-surface error $e(x_{1,i})$ at the positions $x_{1,i}$ after the beam-level correction.

\subsubsection{Effect on the footprint-level correction}

Because the footprint-level correction determines the corrected position from the cost of a single footprint alone, it is directly affected by the reference-surface error. As Eq. (4.9) shows, the cost contains a reference-surface error that differs from candidate to candidate within the search disc. Moreover, in grid search, candidates close to the observed ground elevation line up along a contour line of the reference surface, so a slight error changes which candidate is selected. For this reason, it is difficult to relate the magnitude of the reference-surface error to the magnitude of the deviation of the corrected position.

The residual obtained by evaluating, on the high-resolution DTM, the position $x_2$ corrected with the super-resolved reference surface can be decomposed as $z_{obs}-R_{DEM1A}(x_2)=e(x_2)+\left(z_{obs}-R(x_2)\right)$. The second term on the right-hand side is the residual that the correction minimizes; when grid search selects a candidate that nearly matches the observed ground elevation, this term is nearly 0, and the evaluated residual is nearly equal to $e(x_2)$. In this case, the residual measures not the correctness of the correction but the reference-surface error at the corrected position. In Section 5.4.1, whether the footprint-level corrected positions from the super-resolved reference DTM agree with those from the high-resolution DTM is examined directly.

\section{Results}

Using the super-resolution models and the geolocation correction methods described in the Methods section, we corrected the geolocation of GEDI data in western Japan. We first describe the dataset created for training and then report the super-resolution and geolocation correction results for the test blocks.

\subsection{Experimental setup}

\subsubsection{Dataset}

This study targets forested areas of Japan at latitudes of about 30$^{\circ}$ to 36$^{\circ}$ and longitudes of about 128$^{\circ}$ to 135$^{\circ}$. The unit of data used for training and evaluation is a patch of 20 $\times$ 20 arcseconds in latitude and longitude, which corresponds to 500$\times$500 pixels on the 1 m resolution digital terrain model (DEM1A), 50$\times$50 pixels on the 10 m resolution digital terrain model (DEM10B), and 21$\times$21 sample points on FABDEM, including those on the patch edges.

The test area is defined as ten blocks delimited by coordinates. Each block spans 20 arcminutes in latitude by 30 arcminutes in longitude and was selected by the authors so as to include areas with extensive forests and mountains. A buffer of 5 arcminutes ($\approx$ 9 km) in each of the latitude and longitude directions is placed around each test block, and the area within the buffer is used for neither training nor validation. The area that belongs to neither the test area nor the buffers is split into training and validation areas by random splitting in units of secondary meshes. The fraction of secondary meshes assigned to validation is 15\%.

The patches used for training and validation are placed inside tertiary-mesh tiles with a stride of 200 pixels, whereas the patches of the test blocks are placed at 400-pixel intervals on an absolute grid so that they cover the whole block without gaps. In both cases, the patch edges are placed on 10 m pixel boundaries and at integer arcseconds (details in Appendix B).

Of the extracted patches, those with at least 90\% valid DEM1A pixels, 100\% valid DEM10B pixels, and at least 95\% valid FABDEM pixels were used for training and evaluation. However, patches in which non-natural classes such as urban areas and water bodies occupy at least 80\% of the area in HRLULC were excluded. This procedure yields 261,294 training patches ($\approx$ 25,700 $\mathrm{km}^2$), 46,164 validation patches ($\approx$ 4,500 $\mathrm{km}^2$), and 67,170 test patches in total ($\approx$ 11,000 $\mathrm{km}^2$) (Fig. 3.2, Table 5.1).

\begin{table}[t]
\centering
\small
\caption{Numbers of patches and areas of the training, validation, and test splits. The areas are approximate.}
\label{tab:split}
\begin{adjustbox}{max width=\columnwidth}
\begin{tabular}{lrr}
\toprule
Split & Patches & Ground area (km$^{2}$) \\
\midrule
train & 261,294 & 25,663 \\
valid & 46,164 & 4,522 \\
test & 67,170 & 10,959 \\
\textbf{Total} & 374,628 & 41,145 \\
\bottomrule
\end{tabular}
\end{adjustbox}
\end{table}

\subsubsection{Reference DTMs compared}

The 10 m resolution DTM is super-resolved by the super-resolution models and evaluated against the 1 m resolution DTM. Geolocation correction is also performed with the generated reference surfaces, and its results are evaluated. The five reference surfaces compared are generated as follows.

\begin{labeledlist}
\item[(M0):] A reference surface obtained by bilinear interpolation of DEM10B onto the 1 m resolution grid.
\item[(M1):] A reference surface generated by the super-resolution model of Section 4.1 with DEM10B as the primary input and FABDEM as the auxiliary DTM.
\item[(M2):] A reference surface generated by a model trained in the same way as (M1) but without the FABDEM input. FABDEM is not used in validation or testing either.
\item[(M3):] A model trained in the same way as (M1) but with a synthetic DTM created from DEM1A, instead of DEM10B, as input. The input in validation and testing is also synthetic data.
\item[(M4):] A model trained in the same way as (M3) but without the FABDEM input. FABDEM is not used in validation or testing either.
\end{labeledlist}
The input data differ among them. In particular, the models without the FABDEM input do not perform mid-fusion and have no encoder for FABDEM. The synthetic DTM of (M3) and (M4) is a 10 m grid surface obtained by averaging DEM1A over blocks of 10$\times$10 pixels; the details are given in Appendix B. Hereafter, the notation (M0) to (M4) refers not only to the bilinear interpolation or the super-resolution model itself but also to the generated reference surface and to the results of geolocation correction with that reference surface.

\subsubsection{Training}

We trained the super-resolution models with the training patches. Adam was used for optimization, and MultiStepLR was applied as the learning-rate scheduler. The initial learning rate was $1 \times 10^{-4}$, and it was halved at epochs 40, 60, 80, and 90. The batch size was 20, the maximum number of epochs was 100, and early stopping was not used.

During training, one of eight transformations, consisting of rotations by multiples of 90$^{\circ}$ and flips about the vertical or horizontal axis, was applied at random to each input pair. This was done to reduce the strong directional bias of the mountains in the study area. The model that minimized the RMSE on the validation patches was taken as the best model.

\subsubsection{Evaluation metrics}

The accuracy of super-resolution is evaluated with elevations in meters on the valid DEM1A pixels of the test patches. For the elevation error, that is, the difference obtained by subtracting the DEM1A value from the value of the generated surface, we show the RMSE and the MAE over all pixels of all test patches, the mean of the difference (Avg. Bias), and the standard deviation across patches of the per-patch mean difference (SD Bias). SD Bias represents the size of the elevation deviations that vary from patch to patch. The reproduction of terrain shape is evaluated by RMSE-Slope and RMSE-Aspect as defined by \textcite{Zhang2022-uq}. The improvement over bilinear interpolation is expressed as the percentage reduction in the RMSE. Table 5.2 shows the results of super-resolving the 10 m resolution DTM in the test area and comparing it with DEM1A. Here, the results of bilinear interpolation of DEM10B and of the synthetic data are also included as baselines.

Next, reference surfaces of the test area were created with the super-resolution models and bilinear interpolation, and geolocation correction was performed with these reference surfaces. The correction targets are the 333,902 footprints belonging to the 2,129 beams in the ten test blocks. The same footprints were also corrected with DEM1A as the reference DTM, and this correction is used as the benchmark for the correction results.

In the evaluation by elevation, the Gaussian-weighted elevation of DEM1A is taken at the corrected position, and its residual from the ground elevation observed by GEDI is expressed by the MAE, RMSE, and Bias. In the evaluation by canopy height, GEDI RH98 is compared with the maximum of the DCHM within a radius of 12.5 m, and the comparison is expressed by the MAE, RMSE, Bias, and Pearson correlation coefficient. In both cases, Bias is the mean of the difference obtained by subtracting GEDI from the reference data.

Tables 5.3 and 5.4 show the results of the geolocation correction. Table 5.3 shows the results evaluated by elevation, and Table 5.4 those evaluated by canopy height. The evaluation by canopy height is performed only in five of the ten test blocks (5334\_SE, 5234\_NW, 5033\_NE, 5033\_SE, and 4932\_NE). In Tables 5.3 and 5.4, for each reference DTM, the evaluated footprints are the intersection of the footprints that remained evaluated in the three corrections, that is, the beam-level correction and the two footprint-level methods. The results without correction on the same evaluated footprints are also shown. The numbers of evaluated footprints in Tables 5.3 and 5.4 are those after removing correction failures; in Table 5.3 they correspond to 58.3\% to 69.3\% of all correction targets, the smallest being for (M0) and the largest for DEM1A. Most of the removed footprints are those whose solution reached the rim of the search disc in the footprint-level grid search. In Table 5.3, only the beam-level rows have up to three fewer evaluated footprints than the other stages, because for some footprints the Gaussian-weighted elevation of DEM1A could not be obtained at the position after the beam-level correction.

\begin{table*}[t]
\centering
\small
\setlength{\tabcolsep}{3.5pt}
\caption{Results of evaluating the reference surfaces generated by the super-resolution models and the elevation surfaces created by bilinear interpolation on all 67,170 test patches. Because interpolating DEM10B, which is real data, and interpolating the synthetic data give very different results, two results are given for bilinear interpolation.}
\begin{adjustbox}{max width=\textwidth}
\begin{tabular}{llrrrrrr}
\toprule
Model & Input & RMSE (m) & MAE (m) & Avg. Bias (m) & SD Bias (m) & RMSE-Slope ($^{\circ}$) & RMSE-Aspect ($^{\circ}$) \\
\midrule
Bilinear (M0) & DEM10B & 5.70 & 4.22 & 1.476 & 2.668 & 11.4 & 45.8 \\
\textbf{M1} & \textbf{DEM10B + FABDEM} & \textbf{3.81} & \textbf{2.82} & \textbf{0.162} & \textbf{1.642} & \textbf{9.6} & \textbf{41.4} \\
M2 DEM10B only & DEM10B & 5.32 & 3.93 & -0.008 & 2.650 & 10.2 & 44.1 \\
Bilinear & Synthetic & 0.95 & 0.63 & 0.005 & 0.050 & 7.9 & 33.6 \\
M3 synthetic & Synthetic + FABDEM & 0.42 & 0.27 & -0.001 & 0.003 & 5.9 & 27.0 \\
M4 synthetic only & Synthetic & 0.42 & 0.27 & 0.005 & 0.003 & 5.9 & 27.0 \\
\bottomrule
\end{tabular}
\end{adjustbox}
\end{table*}
\begin{table*}[t]
\centering
\small
\caption{Results of evaluating the geolocation correction by elevation for the target footprints in the test area.}
\begin{adjustbox}{max width=\textwidth}
\begin{tabular}{llrrrr}
\toprule
Reference DTM & Correction stage & Evaluated & MAE (m) & RMSE (m) & Bias (m) \\
\midrule
M0 & No correction & 194,536 & 5.02 & 6.72 & -1.47 \\
 & beam-level & 194,535 & 4.31 & 5.79 & -1.46 \\
 & beam-level+ footprint-level(gs) & 194,536 & 3.88 & 5.11 & -1.64 \\
 & beam-level+ footprint-level(fa) & 194,536 & 3.79 & 5.02 & -1.57 \\
\midrule
M1 & No correction & 213,558 & 4.86 & 6.49 & -0.93 \\
 & beam-level & 213,558 & 3.92 & 5.24 & -0.85 \\
 & beam-level+ footprint-level(gs) & 213,558 & 2.58 & 3.40 & -0.31 \\
 & beam-level+ footprint-level(fa) & 213,558 & 2.61 & 3.45 & -0.34 \\
\midrule
M2 & No correction & 196,825 & 4.83 & 6.46 & -0.89 \\
 & beam-level & 196,824 & 4.09 & 5.48 & -0.81 \\
 & beam-level+ footprint-level(gs) & 196,825 & 3.55 & 4.70 & -0.29 \\
 & beam-level+ footprint-level(fa) & 196,825 & 3.46 & 4.59 & -0.32 \\
\midrule
M3 & No correction & 231,249 & 4.72 & 6.28 & -0.73 \\
 & beam-level & 231,247 & 3.72 & 4.94 & -0.66 \\
 & beam-level+ footprint-level(gs) & 231,249 & 0.09 & 0.52 & -0.04 \\
 & beam-level+ footprint-level(fa) & 231,249 & 0.93 & 1.70 & -0.08 \\
\midrule
M4 & No correction & 231,186 & 4.71 & 6.28 & -0.73 \\
 & beam-level & 231,183 & 3.71 & 4.94 & -0.66 \\
 & beam-level+ footprint-level(gs) & 231,186 & 0.09 & 0.52 & -0.04 \\
 & beam-level+ footprint-level(fa) & 231,186 & 0.93 & 1.70 & -0.09 \\
\midrule
DEM1A & No correction & 231,473 & 4.73 & 6.29 & -0.73 \\
 & beam-level & 231,471 & 3.72 & 4.95 & -0.66 \\
 & beam-level+ footprint-level(gs) & 231,473 & 0.05 & 0.51 & -0.04 \\
 & beam-level+ footprint-level(fa) & 231,473 & 0.93 & 1.70 & -0.08 \\
\bottomrule
\end{tabular}
\end{adjustbox}
\end{table*}
\begin{table*}[t]
\centering
\small
\caption{Results of evaluating the geolocation correction by canopy height for the target footprints in the test area.}
\begin{adjustbox}{max width=\textwidth}
\begin{tabular}{llrrrrr}
\toprule
Reference DTM & Correction stage & Evaluated & MAE (m) & RMSE (m) & Bias (m) & Pearson $r$ \\
\midrule
M0 & No correction & 57,749 & 5.46 & 7.51 & -0.64 & 0.500 \\
 & beam-level & 57,745 & 5.37 & 7.41 & -0.64 & 0.516 \\
 & beam-level+ footprint-level(gs) & 57,742 & 5.35 & 7.39 & -0.68 & 0.521 \\
 & beam-level+ footprint-level(fa) & 57,747 & 5.27 & 7.32 & -0.71 & 0.533 \\
\midrule
M1 & No correction & 62,602 & 5.44 & 7.49 & -1.02 & 0.499 \\
 & beam-level & 62,603 & 5.34 & 7.37 & -1.01 & 0.517 \\
 & beam-level+ footprint-level(gs) & 62,596 & 5.26 & 7.32 & -1.20 & 0.529 \\
 & beam-level+ footprint-level(fa) & 62,607 & 5.20 & 7.26 & -1.19 & 0.540 \\
\midrule
M2 & No correction & 58,048 & 5.38 & 7.39 & -1.09 & 0.521 \\
 & beam-level & 58,041 & 5.30 & 7.29 & -1.11 & 0.537 \\
 & beam-level+ footprint-level(gs) & 58,045 & 5.24 & 7.25 & -1.26 & 0.547 \\
 & beam-level+ footprint-level(fa) & 58,061 & 5.17 & 7.18 & -1.26 & 0.558 \\
\midrule
M3 & No correction & 66,482 & 5.34 & 7.36 & -1.08 & 0.539 \\
 & beam-level & 66,481 & 5.23 & 7.22 & -1.08 & 0.559 \\
 & beam-level+ footprint-level(gs) & 66,472 & 5.03 & 7.05 & -1.28 & 0.586 \\
 & beam-level+ footprint-level(fa) & 66,490 & 4.98 & 7.02 & -1.29 & 0.592 \\
\midrule
M4 & No correction & 66,565 & 5.34 & 7.34 & -1.07 & 0.541 \\
 & beam-level & 66,567 & 5.23 & 7.20 & -1.08 & 0.560 \\
 & beam-level+ footprint-level(gs) & 66,565 & 5.04 & 7.06 & -1.27 & 0.584 \\
 & beam-level+ footprint-level(fa) & 66,565 & 4.99 & 7.00 & -1.28 & 0.594 \\
\midrule
DEM1A & No correction & 66,535 & 5.35 & 7.36 & -1.07 & 0.541 \\
 & beam-level & 66,543 & 5.24 & 7.23 & -1.07 & 0.560 \\
 & beam-level+ footprint-level(gs) & 66,529 & 5.05 & 7.09 & -1.26 & 0.584 \\
 & beam-level+ footprint-level(fa) & 66,544 & 4.99 & 7.03 & -1.28 & 0.593 \\
\bottomrule
\end{tabular}
\end{adjustbox}
\end{table*}

\subsection{Super-resolution accuracy}

\subsubsection{Overall accuracy}

According to Table 5.2, the reference surface generated by (M1) reduced the RMSE by 33.1\% relative to bilinear interpolation, from 5.70 m to 3.81 m. For (M2), which does not take FABDEM as input, the RMSE decreased only to 5.32 m. Even a reference DTM obtained as a simple weighted average of DEM10B and FABDEM, without applying deep learning, has an RMSE of 4.96 m (Appendix C), and the accuracy of (M1) is improved both by inputting the auxiliary DTM and by fusing it inside the super-resolution model. The mean of the bias is 0.16 m for (M1) and $-0.01$ m for (M2), compared with 1.48 m for bilinear interpolation, and both are close to 0. In contrast, the standard deviation of the bias is 2.65 m for (M2), almost unchanged from the 2.67 m of bilinear interpolation, whereas it is 1.64 m for (M1), 38\% smaller. This can be interpreted as follows: with the primary input of this study (DEM10B) alone, the model can learn to remove a systematic error that is constant over the whole area, but an auxiliary input whose origin differs from that of the primary input is needed to greatly reduce the elevation deviations that vary from place to place and the errors of individual patches.

For (M3) and (M4), which use synthetic data, the error with respect to the ground-truth data is extremely small, with an RMSE below 1 m even for bilinear interpolation. (M3) and (M4) can further halve the RMSE, which shows the effectiveness of the super-resolution model itself. In both cases, the bias is extremely small, and there is almost no difference in accuracy between (M3) and (M4), which differ only in whether FABDEM is used. We consider that FABDEM does not contribute to improving the accuracy because no bias arises when the synthetic data are created.

RMSE-Slope and RMSE-Aspect are indices of how well the terrain features are preserved in the generated surface \parencite{Zhang2022-uq}. For (M1), both decreased slightly from those of bilinear interpolation, but they remain much larger than the 5.9$^{\circ}$ and 27.0$^{\circ}$ of (M3), which was trained on synthetic data. Because slope and aspect are derived from the elevation differences between neighboring pixels, we consider that the improvement is small for (M1), which does not generate fine relief, as shown in Section 5.2.2.

\subsubsection{Terrain roughness and accuracy}

What the super-resolution model of this study, which adopts the Global skip, learns is the fine relief that does not appear in the input DTM of 10 m resolution, and we consider that super-resolution is more difficult for terrain with more of this relief. We therefore express the amount of fine relief by the $\mathrm{SD_{restopo}}$ of Section 3.2.4 and examine how much fine relief the generated reference surfaces have.

Fig. 5.1 shows the density distributions of $\mathrm{SD_{restopo}}$ over all test patches for each reference surface. All the density distributions are unimodal, so the positions of the distributions allow a rough comparison of the amount of relief among the reference surfaces. The distribution of (M1) lies at smaller values not only than that of DEM1A but also than that of (M0), the bilinear interpolation of DEM10B. In contrast, the distribution of (M3), which was trained on synthetic data, lies between those of the bilinear interpolation of the synthetic data and of DEM1A. This can be interpreted as follows: (M1) outputs a smooth surface without generating relief that cannot be predicted from the input, whereas (M3), which receives more information as input, generates finer relief. We consider that this is because we selected the super-resolution model that prioritizes the numerical accuracy of elevation. For relief that is not determined by the input, a model trained with a per-pixel error as the loss outputs a smooth surface in which the possible relief is averaged pixel by pixel \parencite{Ledig2017-ft}.

Fig. 5.2 shows examples for three test patches with different $\mathrm{SD_{restopo}}$. The output of (M1) roughly matches DEM1A in the arrangement of valleys and ridges but does not reproduce the fine relief of DEM1A. Fig. 5.3 visualizes the residual $|r_i-\mu_r|$ of each generated pixel from the smoothed surface. In the surface generated by bilinear interpolation of DEM10B, the seams between the 10 m resolution pixels become creases on the generated surface, and residuals exist only on the grid lines. The residuals of (M1) are smaller than those of bilinear interpolation and vary smoothly. In (M3), the residuals remain smooth but their values are larger.

Fig. 5.4 shows the accuracy when the patches are binned by the $\mathrm{SD_{restopo}}$ of DEM1A. In the range from 0.2 m to 0.8 m, to which most DEM1A patches belong, (M1) reduced the RMSE by 30\% to 36\% relative to bilinear interpolation, and in the range where most of the data lie, the change in this improvement rate due to terrain roughness is small. On the other hand, the RMSE itself is larger for patches with more relief; for (M1), it increases from 2.61 m in the 0.2--0.3 m bin to 4.92 m in the 0.7--0.8 m bin.

\begin{figure}[t]
\centering
\includegraphics[width=\columnwidth]{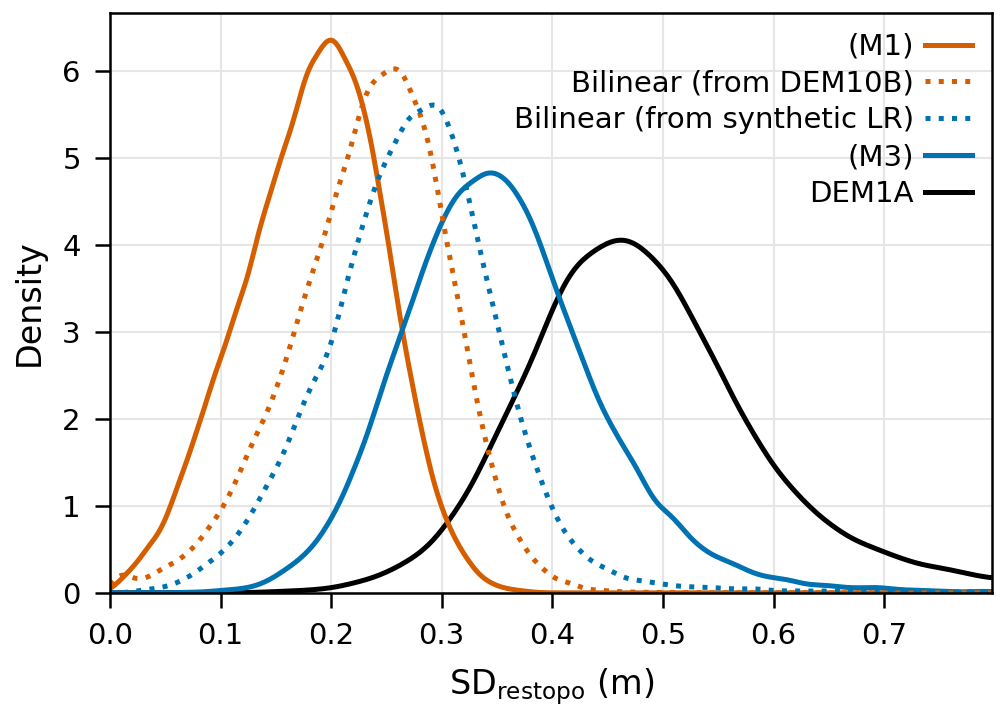}\par\smallskip
\caption{Density distributions of $\mathrm{SD_{restopo}}$ of the reference surfaces generated over the test area. For bilinear interpolation, two results are shown: the interpolation of DEM10B and the interpolation of the synthetic data.}
\label{fig:34b64805-b635-4193-8160-8a6a92fdf833_5602d6b8}
\end{figure}
\begin{figure*}[t]
\centering
\includegraphics[width=\textwidth]{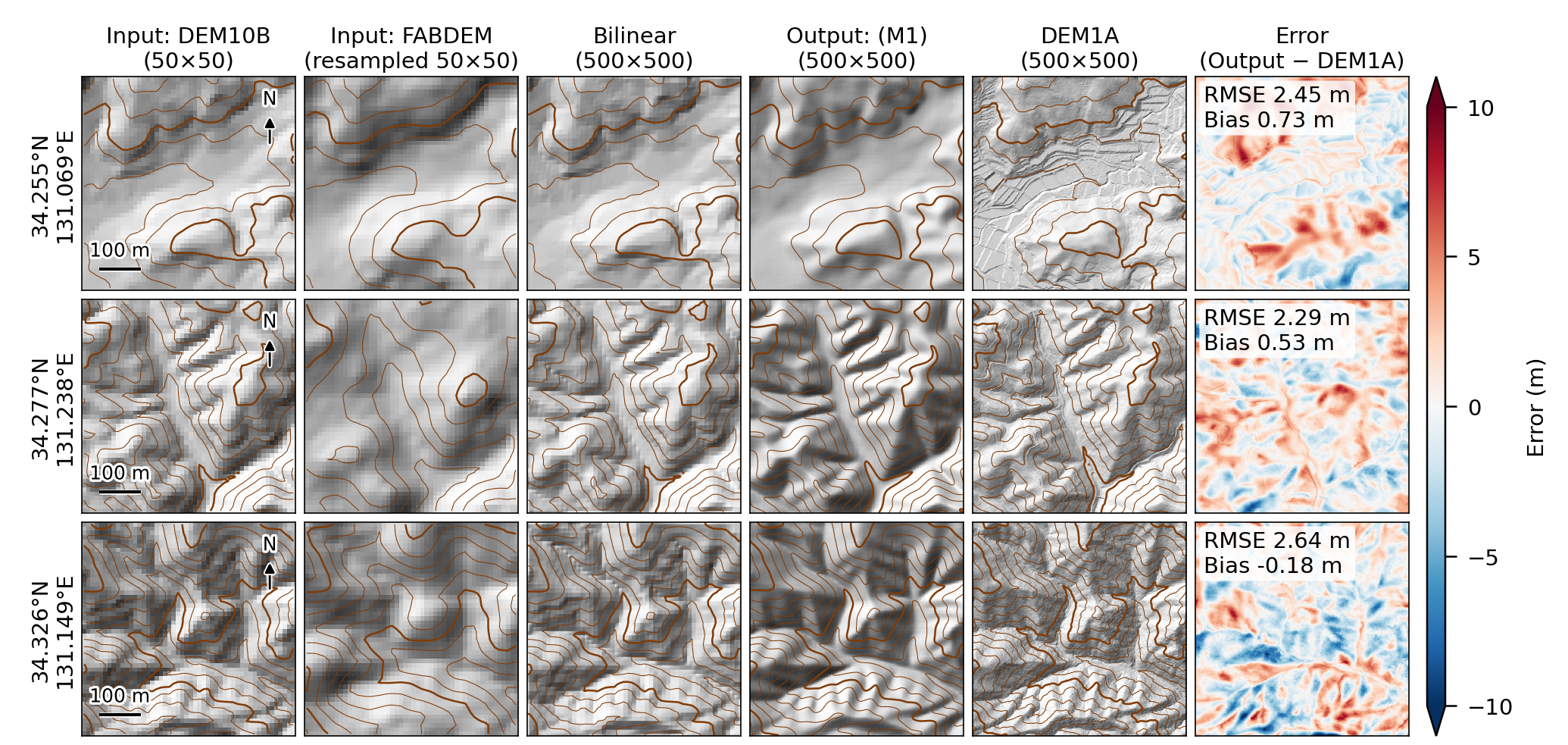}\par\smallskip
\caption{Examples of super-resolution by (M1) for three test patches. From left to right, the columns are the inputs DEM10B and FABDEM, the bilinear interpolation of DEM10B, the output of (M1), DEM1A, and the error obtained by subtracting DEM1A from the output of (M1); the elevation columns are shown as hillshades overlaid with contour lines at 20 m intervals. The three patches are those whose DEM1A $\mathrm{SD_{restopo}}$ is at about the 16th, 49th, and 86th percentiles of all test patches, arranged from top to bottom from the smoothest to the roughest terrain. The RMSE and the bias of each patch are shown on its error panel. The FABDEM column is adapted from FABDEM V1-2 \parencite{Neal2023-fd}, and the (M1) output and error columns are derived from it; these columns are licensed under CC BY-NC-SA 4.0.}
\label{fig:209245b0-c8c4-4cd6-8f7e-00137dddc3e5_42ce0fdd}
\end{figure*}
\begin{figure*}[t]
\centering
\includegraphics[width=\textwidth]{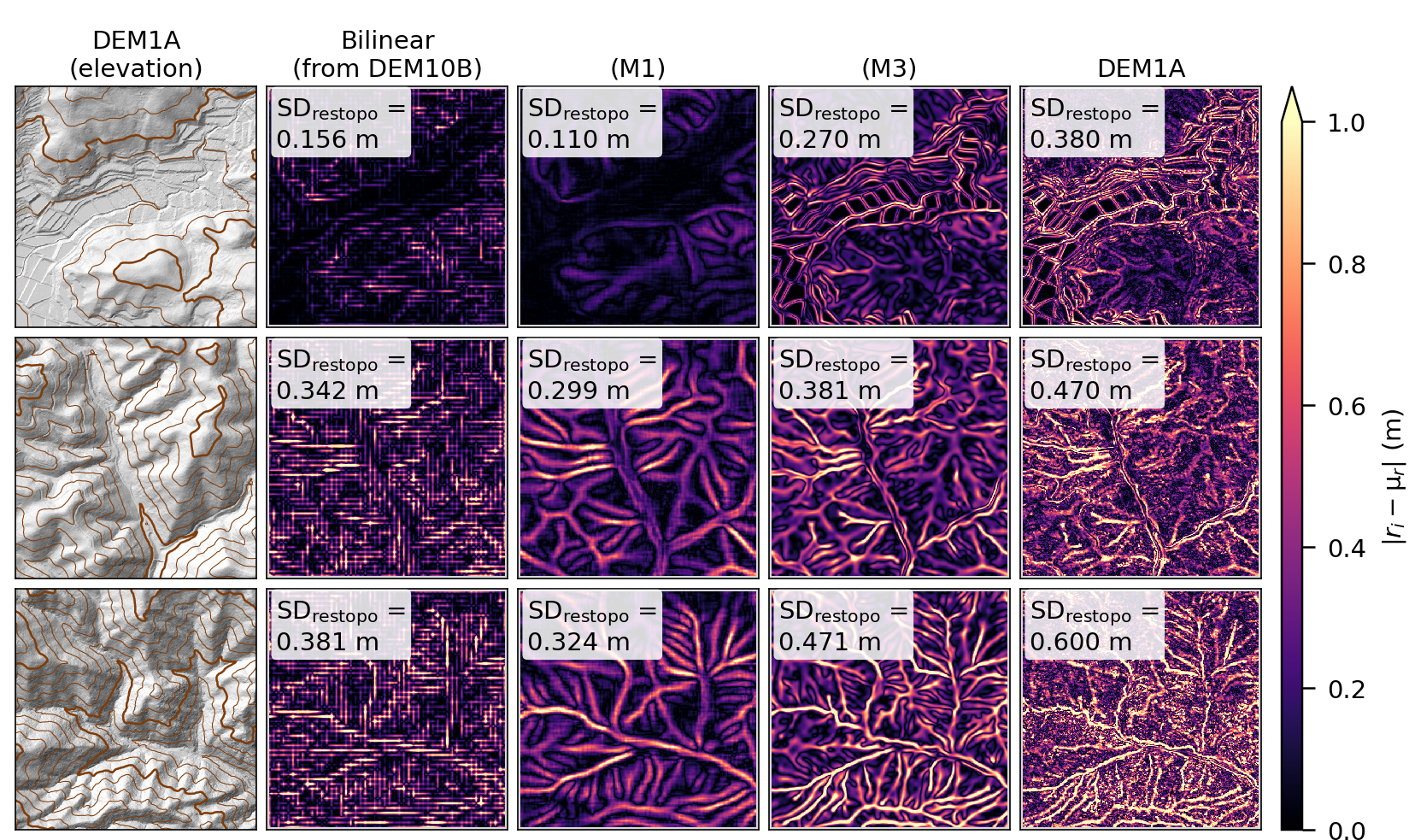}\par\smallskip
\caption{$|r_i-\mu_r|$ of each pixel for the same patches as in Fig. 5.2. The leftmost column shows the hillshade and contour lines of DEM1A as a reference for the terrain. The number at the upper left of each panel is the $\mathrm{SD_{restopo}}$ of that panel. The (M1) and (M3) columns are derived from FABDEM V1-2 \parencite{Neal2023-fd} and are licensed under CC BY-NC-SA 4.0.}
\label{fig:9b8b0505-e443-49e9-82c9-2074c2c7ce98_d8ef55dd}
\end{figure*}
\begin{figure*}[t]
\centering
\includegraphics[width=\textwidth]{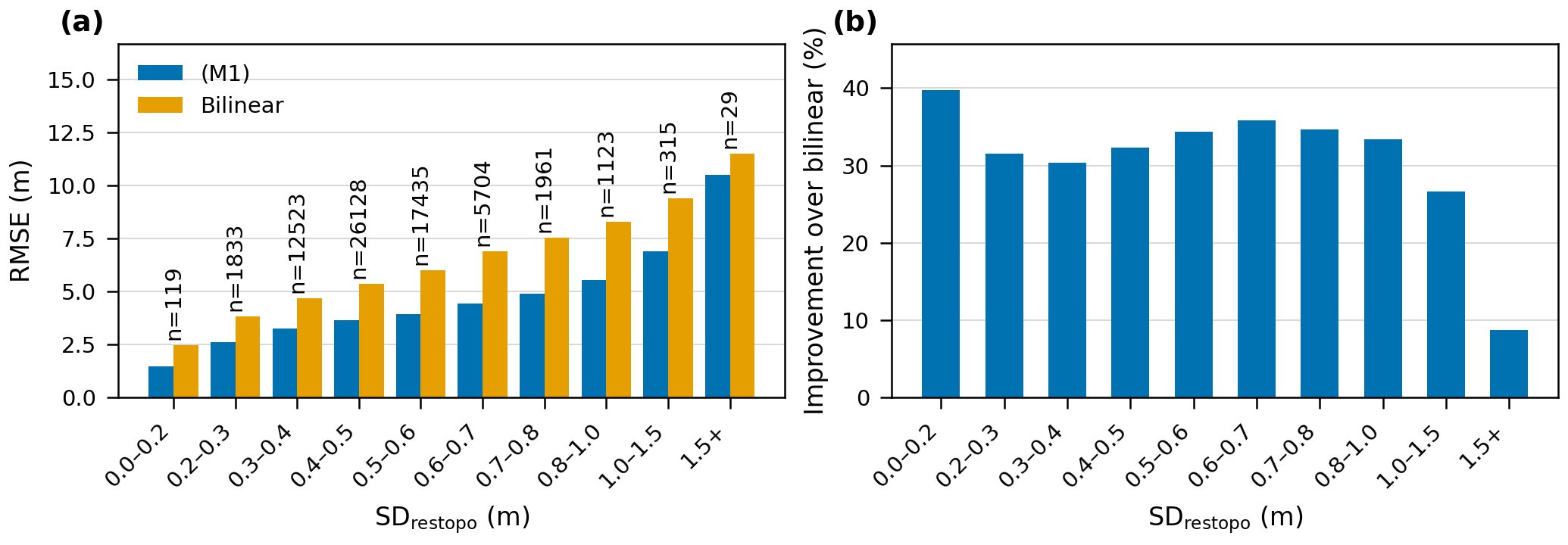}\par\smallskip
\caption{Super-resolution accuracy in each bin when all test patches are binned by the $\mathrm{SD_{restopo}}$ of DEM1A. (a) RMSE of (M1) and of the bilinear interpolation of DEM10B (M0). The numbers above the bars are the numbers of patches in each bin. The RMSE was computed from all pixels of the patches in each bin. (b) Improvement rate of the RMSE of (M1) over bilinear interpolation.}
\label{fig:b1a86206-1d91-4eaf-8cbd-16f5ce92fa23_3fc069f5}
\end{figure*}

\subsubsection{Error spectrum}

The error of super-resolution includes errors due to the fine relief generated by super-resolution and errors at scales larger than the resolution of the input DTM. By computing the error spectrum and examining the component at each wavelength, we check which of the two is dominant as a source of the error. The computation uses the 64,848 test patches in which all DEM1A pixels are valid. Let $N$ be the number of pixels along one side of a patch, $p$ and $q$ the row and column indices of a pixel in the patch, and $\Delta z(p,q)$ the error obtained by subtracting the DEM1A elevation from the elevation of the generated surface. To suppress the effect of the values being cut off at the patch edges, $\Delta z(p,q)$ is multiplied by a Hann window $h(p,q)$, and the two-dimensional discrete Fourier transform given by the following equation is computed.

\begin{equation}
\adjustbox{max width=\linewidth}{$\displaystyle F(k,l)=\sum_{p=0}^{N-1}\sum_{q=0}^{N-1}h(p,q)\,\Delta z(p,q)\exp\left[-\frac{2\pi i\,(kp+lq)}{N}\right]$}
\end{equation}
The integers $k$ and $l$ range from $-N/2$ to $N/2-1$. $F(k,l)$ is one component when the error surface is regarded as a superposition of sinusoids of various wavelengths and orientations, and its wavelength is $N/\sqrt{k^2+l^2}$ pixels. Let $S(k,l)$ be the sum of $|F(k,l)|^2$ over all patches. By Parseval's identity, the sum of $S(k,l)$ over all $k$ and $l$ equals $N^2$ times the sum, over all patches, of the squares of the Hann-windowed error. Because the grid spacing of DEM10B is 10 DEM1A pixels, the components with wavelengths of at least 20 pixels can be represented on the DEM10B grid regardless of their orientation. In the following, this wavelength of 20 pixels serves as an indication of the scales that a 10 m resolution DTM can represent.

Fig. 5.5 shows the error spectra of (M0) to (M3), and Table 5.5 shows the fraction of the sum of $S(k,l)$ accounted for by the components with wavelengths of at least each number of pixels. For (M1), the components with wavelengths of at least 20 pixels account for 98.2\%, and for both (M0) and (M2) they account for 99.1\%. The error of (M1) is smaller than that of (M2) in the components with wavelengths of at least 20 pixels, and the two spectra overlap at wavelengths of less than 20 pixels. For (M3), whose input is synthetic data, the components with wavelengths of at least 20 pixels account for 9.9\%, and most of the error lies in the components with wavelengths of less than 20 pixels.

As described in Section 4.2.1, the correction is performed with the reference surface obtained by averaging the reference DTM with a Gaussian window. We therefore also compute the case in which the error surface in Eq. (5.1) is replaced by the reference-surface error. The reference-surface error $e(p,q)$ at the center of pixel $(p,q)$ is $e(x)$ in Eq. (4.8) and is equal to $\Delta z(p,q)$ averaged with the same Gaussian window as the reference surface. The sum of $S(k,l)$ obtained by replacing $\Delta z(p,q)$ with $e(p,q)$ is 87\% to 91\% of the sum before the replacement for (M0), (M1), and (M2), and 3.5\% for (M3) (Table 5.5). That 87\% remains for (M1) means that applying the Gaussian window reduces the RMSE of the per-pixel error by only 7\%.

\begin{figure}[t]
\centering
\includegraphics[width=\columnwidth]{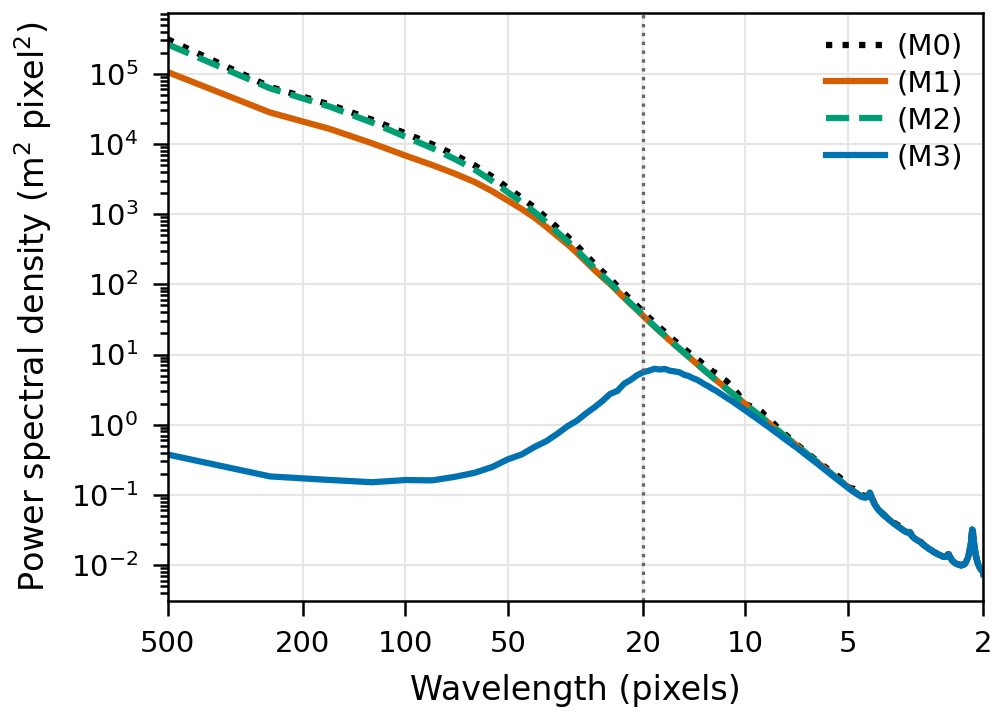}\par\smallskip
\caption{Error spectra of (M0) to (M3) on the test patches. The values are averaged over the components of the same wavelength. The vertical dotted line indicates a wavelength of 20 pixels.}
\label{fig:b3509ab7-764c-435b-9e1b-9e895b8b8b93_b90dfbe1}
\end{figure}
\begin{table*}[t]
\centering
\small
\caption{Fraction (\%) of the sum of $S(k,l)$ accounted for by the components with wavelengths of at least each number of pixels, and the sum of $S(k,l)$ when $\Delta z(p,q)$ is replaced by $e(p,q)$, expressed as a percentage (\%).}
\begin{adjustbox}{max width=\textwidth}
\begin{tabular}{lrrrrr}
\toprule
Reference DTM & $\geq 100$ pixels & $\geq 50$ pixels & $\geq 20$ pixels & $\geq 10$ pixels & $e(p,q)$ \\
\midrule
M0 & 73.8 & 91.4 & 99.1 & 99.8 & 91.2 \\
M1 & 63.1 & 85.1 & 98.2 & 99.5 & 86.6 \\
M2 & 73.6 & 91.4 & 99.1 & 99.8 & 91.1 \\
M3 & 0.0 & 0.2 & 9.9 & 62.9 & 3.5 \\
\bottomrule
\end{tabular}
\end{adjustbox}
\end{table*}

\subsection{Accuracy of geolocation correction}

In the evaluation by elevation in Table 5.3, the ground elevation MAE of (M1) decreased from 4.86 m without correction to 3.92 m after the beam-level correction and to 2.58 m (grid search) and 2.61 m (flow accumulation) after the footprint-level correction. With DEM1A as the reference DTM, it changed from 4.73 m to 3.72 m (beam-level), 0.05 m (grid search), and 0.93 m (flow accumulation). The residual of 2.58 m that remains for (M1) after the correction by the footprint-level grid search is, as described in Section 4.3.3, mainly attributable to the reference-surface error at the corrected position. The same applies to (M0) and (M2).

In Table 5.4, which shows the evaluation by canopy height, the MAE after the footprint-level correction (flow accumulation) is 5.20 m for (M1) and 4.99 m for DEM1A, a difference of 0.21 m. The difference in MAE between (M1) and (M2) is small: on the evaluated footprints of each reference DTM, the MAE of (M2) is 0.03 m smaller, but on the intersection of the evaluated footprints of (M1), (M2), and DEM1A (44,189 footprints), the MAEs after the footprint-level correction (flow accumulation) are 4.92 m, 4.95 m, and 4.82 m, respectively. The correction results of (M3), (M4), and DEM1A were almost the same in both elevation and canopy height.

Looking at each correction method, the elevation residuals in Table 5.3 decrease in the order of no correction, beam-level, and footprint-level for every reference DTM. Of the two footprint-level methods, grid search selects the candidate position whose residual from the reference surface used for the correction is the smallest, whereas flow accumulation selects a flow-weighted centroid that does not coincide with any candidate. The residual measured with DEM1A was smaller for grid search when the reference DTM was DEM1A or close to it. The MAEs are 0.05 m and 0.93 m for DEM1A and 0.09 m and 0.93 m for (M3) and (M4), and for the other reference DTMs the difference between the two methods is 0.1 m or less. For canopy height (Table 5.4), flow accumulation gave a smaller MAE than grid search for every reference DTM, and, in particular, with (M1) and DEM1A as the reference DTM, this tendency was consistent in all test blocks (Fig. 5.6). For both (M1) and DEM1A, the difference in the canopy height MAE between the two methods is small, 0.06 m or less, but its direction is the same in all test blocks. Taking this together with the agreement with the DEM1A-based corrected positions (Section 5.4.1), we recommend flow accumulation for the footprint-level correction.

\begin{figure*}[t]
\centering
\includegraphics[width=\textwidth]{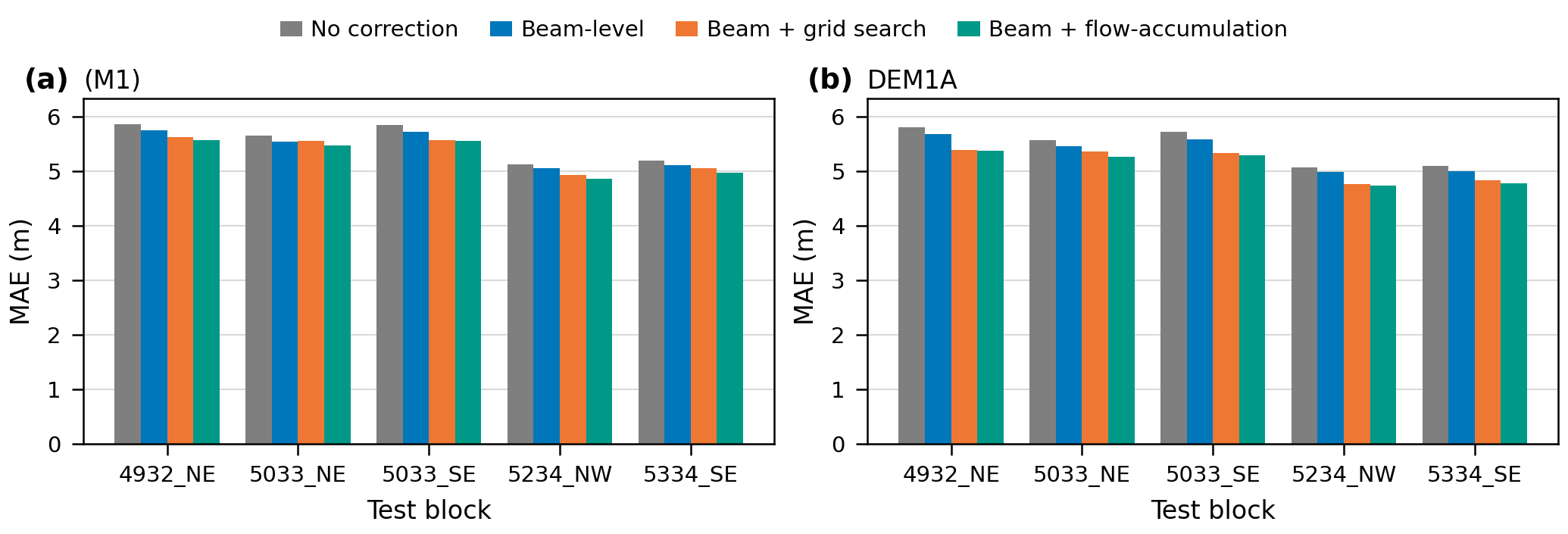}\par\smallskip
\caption{MAE of canopy height in each of the five test blocks where the DCHM is available, for (a) (M1) and (b) DEM1A.}
\label{fig:5e85dcb2-a0ba-4a95-8d87-fad0b23fe88c_ed96e517}
\end{figure*}

\subsection{Reference-surface error and corrected positions}

The differences in accuracy between correction stages evaluated by canopy height are small, and even between reference DTMs, the difference in MAE between (M1) and DEM1A is only 0.10 m on the intersection; therefore, whether the reference DTM generated by super-resolution gives the same correction results as DEM1A cannot be judged from the canopy height accuracy alone. In the following, we therefore examine the influence of the reference DTM on the correction results by the agreement with the DEM1A-based corrected positions and by the descriptive statistics of the reference-surface error.

\subsubsection{Agreement with the DEM1A-based corrected positions}

The agreement of corrected positions is measured by the magnitude of the difference in the offset $u$ at the beam level and by the distance between the corrected positions $x_2$ at the footprint level. The agreement of corrected positions measures whether the same results as those of the existing method with DEM1A as the reference DTM are obtained; it does not measure whether the corrected position with the super-resolved reference DTM has moved closer to or farther from the true position. At the footprint level, for each method, the footprints that were evaluated for that method with both the reference DTM being compared and DEM1A are counted. At the beam level, the beams whose beam-level solution did not reach the rim of the search disc with either of them are counted; for (M1), these are 1,736 of the 2,129 beams. Note that, unlike in the preceding results, the evaluated footprints and beams differ from correction to correction.

Fig. 5.7 shows the distributions of the offsets of the geolocation correction with DEM1A (top row) and the (M1) corrected positions relative to the corresponding DEM1A corrected positions (bottom row). The beam-level offset $u$ spreads over the whole search disc, and its median magnitude is 7.7 m (Table 5.6). The footprint-level offset $v$ spreads almost uniformly over the whole search disc with grid search, whereas it concentrates near the origin with flow accumulation (Fig. 5.7(b)(c)). The discrepancies between the DEM1A and (M1) corrections in the bottom row show that, at the beam level, the discrepancies lie within a narrower range than the distribution of the offset $u$, whereas, at the footprint level, the discrepancies of the corrected positions spread over a range larger than the search disc.

\begin{figure}[t]
\centering
\includegraphics[width=\columnwidth]{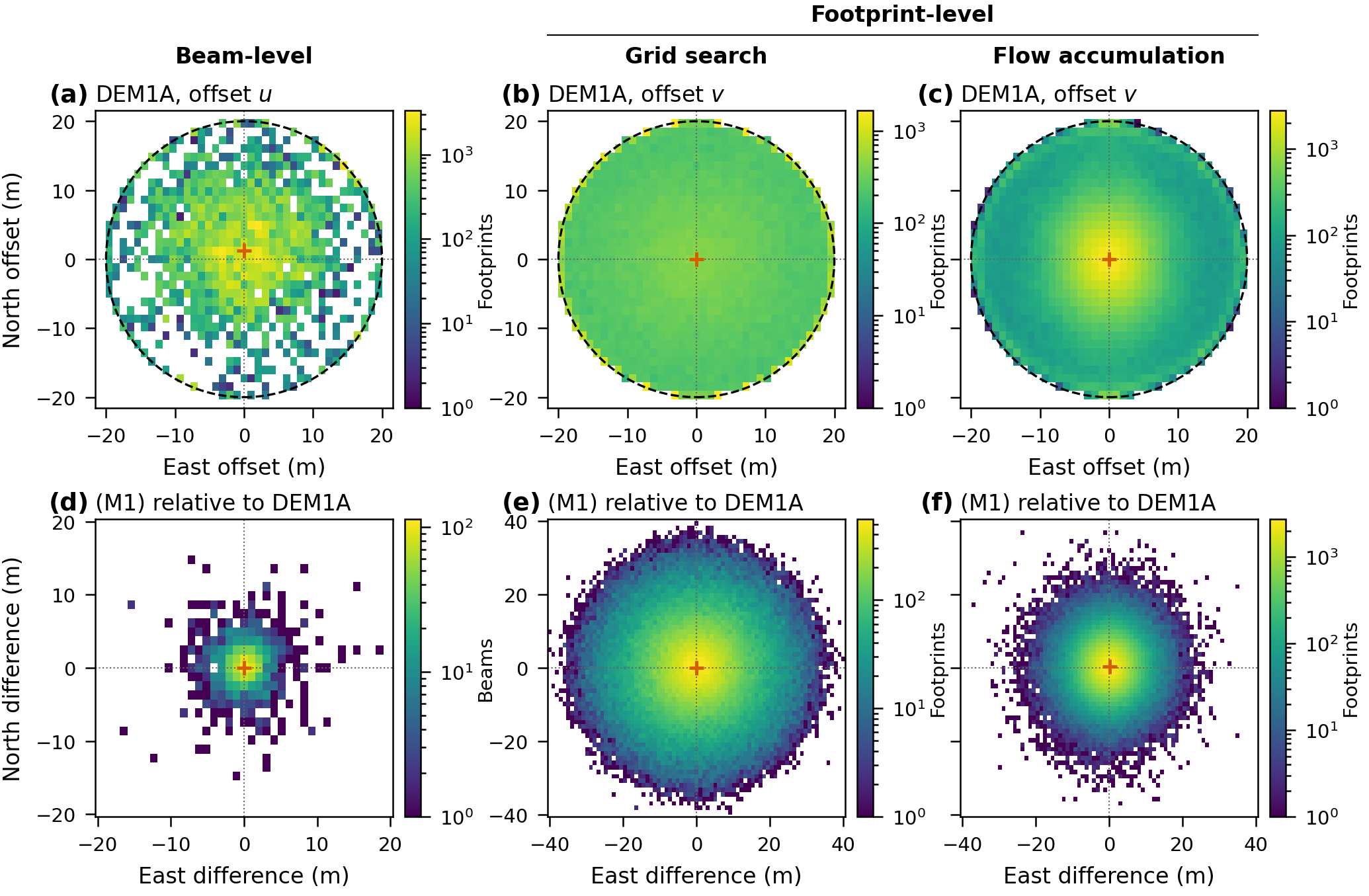}\par\smallskip
\caption{Distributions of the offsets of the geolocation correction. The top row (a)--(c) shows the offsets of the correction with DEM1A: (a) the beam-level offset $u$ (origin at the recorded position $x_0$), and the footprint-level offset $v$ by (b) grid search and (c) flow accumulation (origin at the position $x_1$ after the beam-level correction). The bottom row (d)--(f) shows the (M1) corrected positions relative to the corresponding DEM1A corrected positions, each of which is taken as the origin: (d) beam-level (per beam), and (e) grid search and (f) flow accumulation (per footprint). The dashed lines in the top row show the search disc of radius 20 m. The red + marks the median.}
\label{fig:0cc27801-dcb4-41dc-960a-e5db49e3a8fe_5ebf1689}
\end{figure}
Table 5.6 shows how far the corrected positions with each of the reference DTMs (M0) to (M3) lie from the DEM1A-based corrected positions. At the beam level, the median magnitude of the difference between the offsets of (M1) and DEM1A was 2.4 m, and the difference was within 5 m for 84.6\% of the beams and at least 10 m for 2.8\% of the beams (Fig. 5.7(d)). This median difference is less than one third of the median magnitude of the DEM1A offset (7.7 m). For (M0) and (M2), the median differences were 4.2 m and 4.0 m and the fractions within 5 m were 56.2\% and 60.6\%, respectively, and for (M3), the difference was within 0.5 m for 86.4\% of the beams. At the footprint level, the median distance between the (M1) and DEM1A corrected positions was 11.6 m for grid search and 5.6 m for flow accumulation, and the distance for flow accumulation is about the same as its median displacement (7.4 m). Even with flow accumulation, the fraction within 0.5 m was only 0.7\% and the fraction within 2.5 m only 14.6\%. For (M3), 88.0\% of the corrected positions by flow accumulation lie within 2.5 m, and (M3) almost reproduces the DEM1A-based corrected positions. With grid search, even with DEM1A, the candidates close to the observed ground elevation line up along a contour line of the reference surface, and small errors change which of them is selected as the solution; therefore, the fraction within 0.5 m is low, 20.9\%, even for (M3). The differences between the (M1) and DEM1A corrected positions show no bias toward a fixed direction overall, and the medians of the east--west and north--south components are within 0.4 m of zero at every stage.

In summary, the beam-level correction results of (M1) and (M3) largely agree with those of DEM1A, and the degree of agreement decreases for (M0) and (M2), whose reference-surface errors are larger. The footprint-level correction results agree for flow accumulation with (M3), whose reference-surface error is small, but not for (M0), (M1), and (M2). The degree of agreement follows the order of the magnitude of the reference-surface error (Table 5.9).

The degree of beam-level agreement increases with the number of footprints $n$ in the beam (Table 5.7). The median difference in the offset of (M1) is 4.3 m for beams with $n$ below 60, 2.5 m for $n$ from 60 to below 120, 2.1 m for $n$ from 120 to below 240, and 1.6 m for $n$ of 240 or more, and the fraction within 5 m rises from 56.9\% to 99.1\%. The same tendency holds for (M0) and (M2), and for (M3), whose reference-surface error is small, the median difference is 0 in every range of $n$.

Table 5.8 shows the results when the evaluated footprints of the beam-level correction are extended to all footprints of the beams whose beam-level solution did not reach the rim of the search disc. The evaluated footprints amount to 90.3\% of the correction targets for (M1) and 91.2\% for DEM1A. The ground elevation MAE decreased from 6.28 m to 5.52 m for (M1) and from 6.31 m to 5.47 m for DEM1A.

\begin{table*}[t]
\centering
\small
\setlength{\tabcolsep}{3.5pt}
\caption{Agreement with the DEM1A-based corrected positions. For the per-beam difference in the offset at the beam level and the distance between the corrected positions at the footprint level, the fractions within 0.5 m, within 2.5 m, within 5 m, and at least 10 m, the median, and the number of units are shown. The median displacement of each stage is also given for the reference DTM and for DEM1A.}
\begin{adjustbox}{max width=\textwidth}
\begin{tabular}{llrrrrrrl}
\toprule
Reference DTM & Stage & Units & Within 0.5 m (\%) & Within 2.5 m (\%) & Within 5 m (\%) & At least 10 m (\%) & Median (m) & Displacement (m), reference DTM / DEM1A \\
\midrule
M0 & beam-level & 1,697 & 2.2 & 23.8 & 56.2 & 8.1 & 4.2 & 8.4 / 7.7 \\
 & grid search & 167,083 & 0.2 & 3.1 & 10.3 & 66.4 & 13.3 & 12.7 / 11.7 \\
 & flow accumulation & 244,108 & 0.3 & 7.5 & 25.6 & 35.3 & 7.9 & 8.5 / 6.1 \\
\midrule
M1 & beam-level & 1,736 & 6.5 & 56.4 & 84.6 & 2.8 & 2.4 & 8.0 / 7.7 \\
 & grid search & 186,543 & 0.3 & 4.6 & 14.7 & 58.4 & 11.6 & 12.4 / 11.8 \\
 & flow accumulation & 257,964 & 0.7 & 14.6 & 43.4 & 17.5 & 5.6 & 7.4 / 6.2 \\
\midrule
M2 & beam-level & 1,679 & 1.3 & 26.1 & 60.6 & 7.1 & 4.0 & 8.5 / 7.6 \\
 & grid search & 171,307 & 0.2 & 3.2 & 10.9 & 64.8 & 12.8 & 12.7 / 11.6 \\
 & flow accumulation & 245,909 & 0.4 & 8.6 & 28.7 & 30.9 & 7.4 & 8.2 / 6.1 \\
\midrule
M3 & beam-level & 1,823 & 86.4 & 99.1 & 99.8 & 0.1 & 0.0 & 8.0 / 8.0 \\
 & grid search & 212,884 & 20.9 & 30.1 & 40.6 & 39.4 & 7.2 & 12.0 / 12.0 \\
 & flow accumulation & 279,750 & 38.2 & 88.0 & 97.8 & 0.2 & 0.8 & 6.5 / 6.5 \\
\bottomrule
\end{tabular}
\end{adjustbox}
\end{table*}
\begin{table*}[t]
\centering
\small
\caption{Difference in the beam-level offset between (M1) and DEM1A by range of the number of footprints $n$ in a beam. For the same set of beams as in Table 5.6, the number of beams, the median magnitude of the DEM1A offset, the median difference in the offset, and the fraction of differences within 5 m are shown.}
\begin{adjustbox}{max width=\textwidth}
\begin{tabular}{lrrrr}
\toprule
$n$ & Beams & DEM1A offset (m) & Difference (m) & Within 5 m (\%) \\
\midrule
Below 60 & 473 & 8.7 & 4.3 & 56.9 \\
60 to below 120 & 341 & 8.0 & 2.5 & 86.8 \\
120 to below 240 & 462 & 7.5 & 2.1 & 97.0 \\
240 or more & 460 & 6.3 & 1.6 & 99.1 \\
All & 1,736 & 7.7 & 2.4 & 84.6 \\
\bottomrule
\end{tabular}
\end{adjustbox}
\end{table*}
\begin{table*}[t]
\centering
\small
\setlength{\tabcolsep}{3.5pt}
\caption{Evaluated footprints and accuracy when only the beam-level correction is applied. The fraction is the proportion of the evaluated footprints in the correction targets, and the values for elevation and canopy height are MAEs (m).}
\begin{adjustbox}{max width=\textwidth}
\begin{tabular}{lrrrrrrr}
\toprule
Reference DTM & Beams & Footprints & Fraction (\%) & Elevation, no correction & Elevation, beam-level & Canopy height, no correction & Canopy height, beam-level \\
\midrule
M0 & 1,755 & 301,477 & 90.3 & 6.31 & 5.75 & 6.54 & 6.47 \\
M1 & 1,779 & 301,635 & 90.3 & 6.28 & 5.52 & 6.54 & 6.46 \\
M2 & 1,724 & 300,213 & 89.9 & 6.29 & 5.71 & 6.52 & 6.46 \\
M3 & 1,827 & 304,496 & 91.2 & 6.30 & 5.47 & 6.54 & 6.45 \\
DEM1A & 1,826 & 304,654 & 91.2 & 6.31 & 5.47 & 6.54 & 6.45 \\
\bottomrule
\end{tabular}
\end{adjustbox}
\end{table*}

\subsubsection{Reference-surface error}

Table 5.9 shows the descriptive statistics of the reference-surface error $e(x_1)$ evaluated at the position $x_1$ after the beam-level correction, for those evaluated footprints of Table 5.8 at which the reference-surface error can be obtained. As defined in Eq. (4.8), $e(x)$ is the difference in Gaussian-weighted elevation between the reference DTM and DEM1A; unlike the elevation difference in Table 5.2, it is evaluated only at the footprint positions instead of at all pixels of the test area, and for values weighted by the Gaussian window instead of pixel values.

For (M1), the per-footprint $e(x_1)$ has a mean of 0.31 m, a standard deviation of 3.61 m, and a mean absolute value of 2.75 m. Among the reference DTMs, the standard deviation of $e(x_1)$ is remarkably small for (M3), 0.08 m, and increases in the order of (M1), (M2), and (M0), at 3.61 m, 5.16 m, and 5.32 m. (M0) and (M2) differ almost only in the mean (1.70 m and 0.10 m): the super-resolution with DEM10B alone as input removed the deviation common to the whole area but hardly reduced the errors that vary from place to place. The errors that vary from place to place were reduced by (M1), which adds the auxiliary DTM, and its standard deviation is about 30\% smaller than that of (M2). This is the same relationship as that seen in the mean and the standard deviation of the bias in Table 5.2.

\begin{table*}[t]
\centering
\small
\caption{Descriptive statistics of the reference-surface error $e(x_1)$: per-footprint mean, standard deviation, and mean absolute value.}
\begin{adjustbox}{max width=\textwidth}
\begin{tabular}{lrrrrr}
\toprule
Reference DTM & Beams & Footprints & Mean (m) & Standard deviation (m) & Mean absolute value (m) \\
\midrule
M0 & 1,755 & 301,418 & 1.70 & 5.32 & 4.23 \\
M1 & 1,779 & 301,564 & 0.31 & 3.61 & 2.75 \\
M2 & 1,724 & 300,153 & 0.10 & 5.16 & 3.90 \\
M3 & 1,827 & 304,437 & 0.00 & 0.08 & 0.05 \\
\bottomrule
\end{tabular}
\end{adjustbox}
\end{table*}

\section{Discussion}

The question of this study was whether a reference DTM generated by super-resolution gives the same geolocation correction results as DEM1A, and, if not, why. The Results section shows that the answer differs between the correction stages. In the beam-level correction, (M1) gave almost the same offsets as DEM1A, and the degree of agreement increased with the number of footprints in a beam. For (M2), which was super-resolved without the auxiliary DTM, the degree of agreement was lower. In the footprint-level correction, the corrected positions of (M1) did not reproduce those of DEM1A. Below, we first take the beam-level correction, where agreement was obtained, and discuss how far the mechanism of the agreement can be explained from the results in the Results section (Section 6.1). We then discuss why, with the same reference DTM, the footprint-level result differed from the beam-level result, and the precautions for performing footprint-level correction with a super-resolved reference DTM (Section 6.2). Next, we discuss the possibility that the effect of geolocation correction is hardly reflected in canopy height (Section 6.3). We then describe the scales at which the reference-DTM error lies (Section 6.4) and discuss the implications for the super-resolution of real data (Section 6.5). Finally, we discuss the significance and limitations of this study (Section 6.6).

\subsection{Why beam-level offsets agree}

In the beam-level correction, (M1), generated by super-resolution, gave almost the same offsets as DEM1A. This agreement is not a coincidence. This is because the difference between the two offsets is much smaller than the magnitude of the offsets themselves, even though the reference-surface error around the footprints is too large to be neglected. As described in the Methods section, it is difficult to reason directly about how much the reference-surface error moves the offset of an individual beam, and this study goes no further than the interpretation that matching many footprints at once lowers the sensitivity of the offset to the reference-surface error.

The observations in the Results section are all consistent with this interpretation. The offset difference decreases with the number of footprints in a beam, and for beams with many footprints it approaches the pixel spacing of the reference DTM. Because the offsets agreed regardless of the number of footprints under the synthetic-data condition, where the reference-surface error is almost absent, the offset difference cannot be explained merely by the instability of the solution in beams with few footprints, and we consider that the reference-surface error is involved. The lower degree of agreement for (M0) and (M2), whose errors are larger, can be interpreted in the same way. Therefore, we consider that, in the beam-level correction, matching many footprints at once lowers the sensitivity of the offset to the reference-surface error.

\subsection{Why footprint-level corrections diverge}

In contrast, the reason the generated reference DTM did not reproduce the DEM1A-based corrected positions in the footprint-level correction lies in the relation between the elevation difference that the footprint-level correction tries to resolve and the magnitude of the reference-surface error. The footprint-level correction tries to resolve the elevation difference between the observed ground elevation and the reference DTM that remains after the beam-level correction, using a single footprint alone. As Eq. (4.9) shows, the reference-surface error then changes the elevation to be matched at each candidate by the amount of that error. As the ground-elevation MAE in Table 5.8 shows, the mean absolute value of the elevation difference between the observed ground elevation and DEM1A at the positions after the beam-level correction is 5.5 m to 5.8 m regardless of the reference DTM. By comparison, the mean absolute value of the reference-surface error is 0.05 m for (M3), which is negligible, but 2.75 m for (M1), which reaches half of the elevation difference, and 3.90 m and 4.23 m for (M2) and (M0), which exceed two thirds of it (Table 5.9). The footprint-level corrected positions almost reproduced the DEM1A-based corrected positions with flow accumulation for (M3), whose reference-surface error is negligible, but did not reproduce them for (M0), (M1), and (M2). We consider that the closer the reference-surface error is to the elevation difference to be resolved, the more the corrected position is dominated by the reference-surface error.

From the above, footprint-level correction with a reference DTM whose accuracy is comparable to that of (M1) is not a substitute for footprint-level correction with DEM1A. If footprint-level correction is to be performed with a super-resolved reference DTM, the following points require attention. First, because the corrected position deviates from the DEM1A-based corrected position by about as much as the displacement made by the correction, the same corrected positions as those of the DEM1A-based correction cannot be expected for individual footprints. Second, flow accumulation should be used for the correction. The inferiority of grid search in the canopy-height evaluation is not limited to (M1) but also appears in canopy height in the same way when DEM1A is used as the reference DTM, so its cause is not the reference-surface error but a property of the method. This property is the instability of the solution described in the Methods section. The cost map of a single footprint has multiple near-minimum candidates lying along a contour line of the reference surface, and the solution of grid search, which selects only one point, switches among them with only a slight error. Even under the synthetic-data condition, where the reference-surface error is almost absent, few of the grid-search corrected positions were within 0.5 m of the DEM1A-based corrected positions, whereas flow accumulation almost reproduced the DEM1A-based corrected positions. With a super-resolved reference DTM, the reference-surface error is added to this instability, so the conservative behavior of flow accumulation is all the more important. Third, whether the footprint-level correction makes the position worse than the beam-level corrected position has not been determined in this study. This study evaluated positions by the agreement with the DEM1A-based corrected positions, but the DEM1A-based corrected position is not the true position either. That the true position is unknown is not a limitation specific to this study; directly verifying the correctness of the position requires simulated observations with a known geolocation error \parencite{Roy2021-xp} or validation by independent ground surveys \parencite{Tang2023-kw}, which is outside the scope of this study.

\subsection{Limitations of the evaluation by canopy height}

The reason this study measured the influence of the reference DTM not only by canopy height but also by the agreement with the DEM1A-based corrected positions is that the effect of geolocation correction was hardly reflected in canopy height. Even with DEM1A as the reference DTM, the MAE of canopy height changed only from 5.35 m with no correction to 5.24 m after the beam-level correction and 4.99 m after the footprint-level correction (flow accumulation), and the difference between (M1) and DEM1A is 0.10 m on the intersection of the evaluated footprints. For (M3) with synthetic data, the MAE of canopy height agreed with that of DEM1A within 0.02 m, even though approximately 60\% of the grid-search corrected positions were more than 5 m away from the DEM1A-based corrected positions. These results suggest that canopy height hardly responds to the displacements of a few meters to a little over ten meters made by geolocation correction. A similar insensitivity has been reported in previous studies \parencite{Quiros2021-mu,Tang2023-kw,Schleich2023-xz,Yang2024-mi}. Although direct comparison with most previous studies is not possible because the definitions of canopy height and the target forests differ, when the recorded positions were shifted in eight directions in steps of 5 m up to 20 m, as in \textcite{Quiros2021-mu} and \textcite{Kutchartt2022-lz}, the MAE of canopy height increased only from 5.34 m to at most 5.85 m, whereas the MAE of ground elevation increased from 5.17 m to at most 9.69 m (Appendix E). However, this is the case in which all footprints are shifted uniformly. When the destination was selected for each footprint, the MAE of canopy height dropped to 2.54 m at the positions with the smallest canopy-height error, but the MAE of ground elevation increased to 6.32 m. At the positions with the smallest ground-elevation residual, the MAE of ground elevation dropped to 0.40 m, but the MAE of canopy height remained at 5.05 m (Appendix E). In other words, the MAE of canopy height itself can take small values, but such values are not of the kind obtained by a correction based on ground elevation and may be smaller than the ideal lower bound reachable by correction. If the lower bound reachable by correction is close to the current accuracy, canopy height is insensitive to correction, but this is difficult to judge with the data of this study.

This property holds regardless of the reference DTM. Hence, ``almost the same accuracy in canopy height'' does not so much prove that the super-resolved reference DTM gives the same correction results as DEM1A as show that canopy height lacks the resolving power to discriminate between reference DTMs. What can be clearly stated from canopy height is limited to the direction of accuracy differences that is consistent across all test blocks, such as the ranking of the two footprint-level methods.

Possible causes of the weak effect of correction on canopy height include the definition of the reference canopy height, the fact that on slopes RH98 includes the terrain relief within the footprint and can exceed the tree height \parencite{Park2014-of,Li2024-kz}, the variation of tree heights within a stand, and the vertical error of GEDI itself; this study did not separate them. The ground elevation and canopy-top elevation of GEDI contain vertical errors that arise when they are identified from the received waveform, and these errors become larger under dense canopies and on steep slopes \parencite{Hancock2019-wg,Liu2021-xk,Quiros2021-mu}. DTM-based correction cannot distinguish these vertical errors from geolocation errors, and they degrade the accuracy of the correction \parencite{Adam2020-df,Li2025-oz}. Considering these factors, increasing the agreement of canopy height through correction would require using canopy information in determining the position as well, as in the method that determines the position based on agreement with a canopy-height metric \parencite{Wang2025-bj}.

\subsection{Scale of the reference-DTM errors}

Because the correction uses the reference DTM through the Gaussian window, the shorter the wavelength of an error, the less of it reaches the correction. For (M1), the components with wavelengths of at least 20 pixels, which correspond to the scales that a 10 m-resolution DTM can represent, account for 98.2\% of the total of $S(k,l)$, and even after averaging with the Gaussian window, 87\% of the total of $S(k,l)$ remains (Section 5.2.3). Most of the reference-DTM error that affects the correction lies not in the details of the 1 m grid but at scales that a 10 m-resolution DTM can represent. Therefore, improving the accuracy of geolocation correction requires reducing the error at these scales rather than reproducing details.

The primary input, DEM10B, is a DTM produced from the contour lines of topographic maps; it does not contain the shape of the terrain between contour lines, and its elevation accuracy is stated to be within 5 m. For (M2), which used DEM10B alone as input, the standard deviation of the per-patch bias hardly differed from that of bilinear interpolation. What the training of (M2) could remove was the systematic error common to the whole area, which means that the location-dependent elevation errors cannot be corrected from DEM10B alone. Reducing such errors requires using another high-quality DTM as input or adding an information source whose origin differs from that of the primary input. In this study, to reduce the bias of the DTM generated by super-resolution, we added FABDEM, which has a different origin, as the auxiliary input (Section 4.1.2). For (M1), the standard deviation of the per-patch bias decreased by 38\% from that of (M2), and, accordingly, the reference-surface error became smaller, the beam-level agreement increased, and the elevation residual after the footprint-level correction decreased.

Adding input DTMs of different origins is often subject to constraints. Free DTMs covering the globe are limited to those, like FABDEM, derived by removing vegetation from a 30 m-resolution DSM. On the other hand, optical images are acquired repeatedly over the globe at resolutions of 10 m or finer. In DEM super-resolution, there are studies that restore terrain details using high-resolution aerial photographs as auxiliary inputs \parencite{Argudo2018-kc,Kubade2020-nu,Liu2026-jj}, and in the production of global DTMs, methods that remove vegetation-induced errors of DSMs using vegetation height and cover as predictor variables are used \parencite{Hawker2022-ft}. However, what optical images capture is the surface of the canopy, and terrain details cannot be estimated from aerial photographs in areas covered by vegetation \parencite{Argudo2018-kc}. In contrast, in super-resolution that takes a DSM as input and outputs a DTM, there is an example in which canopy height maps and land use were used as auxiliary inputs in addition to aerial photographs to correct the errors of the DSM caused by vegetation and buildings \parencite{Cai2025-xt}. Thus, even if optical images cannot be a source of information on the terrain beneath the canopy, they can make up for the shortage of input DTMs of different origins as a source of information that explains the errors of the input.

For (M3), which used synthetic data produced by coarsening DEM1A as input, the reference-surface error was far smaller than under the real-data condition, and the corrected positions nearly agreed with those of DEM1A, both at the beam level and at the footprint level with flow accumulation. (M1) and (M3) share the same model architecture and were trained under conditions that differ only in the input. The low-resolution data synthesized from DEM1A are merely a coarsened version of the terrain information in DEM1A and do not contain the differences arising from different measurement methods that are contained in the real DTM pair. The fact that the framework of this study gives nearly the same correction results as DEM1A when the input contains sufficient information means that improving the input DTM leads directly to improving the correction results.

\subsection{Implications for super-resolution}

The super-resolution results of this study also provide findings on the super-resolution of real data. First, the super-resolution accuracy of models trained and evaluated on synthetic data produced by coarsening DEM1A was far higher than that on real data. With synthetic data, the RMSE was less than 1 m even with bilinear interpolation, whereas with real data it is 3.81 m even after super-resolution. It is known that a model trained on synthetic pairs of low- and high-resolution data does not even match interpolation on real data, and that super-resolution trained on synthetic data does not work on real data because real low-resolution data contain effects of sensors and acquisition conditions that interpolation cannot reproduce \parencite{Wu2021-hi,Panangian2024-ec}. In this study, not only the accuracy but also the conclusions on the footprint-level correction differed between synthetic and real data. Second, the super-resolution accuracy on real data was improved by fusing the 30 m-resolution FABDEM, whose origin differs from that of the primary input. The auxiliary inputs in previous studies are optical images finer than the input, and they serve to restore details and to correct biases arising from ground objects \parencite{Argudo2018-kc,Kubade2020-nu,Cai2025-xt}. The auxiliary input in this study is a DTM that is coarser than the input and of a different origin, and it reduced only the location-dependent bias without adding details. This bias was hardly reduced by training with DEM10B alone, and with synthetic data, the accuracy hardly changed with or without the auxiliary input (Section 5.2.1). The finding from DEM fusion that fusing DEMs of different origins reduces error \parencite{Okolie2022-hs} can also hold for deep-learning super-resolution that uses a DTM coarser than the input as the auxiliary input, in the form of compensating for the error that training cannot remove.

\subsection{Significance and limitations}

This study showed that beam-level correction gives nearly the same results as with DEM1A even with a super-resolved DTM. With the super-resolved reference DTM (RMSE 3.81 m), obtained by fusing a contour-derived 10 m-resolution DTM with a global 30 m-resolution DTM of a different origin, the beam-level offset agreed with that of DEM1A within 5 m for 84.6\% of the beams. The accuracy of DTM-based geolocation correction of GEDI has so far been verified with DTMs of 0.5--1 m resolution \parencite{Tang2023-kw,Schleich2023-xz,Li2025-oz,Tsutsumida2026-bv}, but the present results suggest that, even in areas where a DTM of roughly 10 m resolution is available, fusing it with a global DTM and super-resolving it to 1 m resolution may give beam-level corrections that are nearly the same as those obtained with a 1 m-resolution DTM. Under the interpretation in Section 6.1, the smaller the reference-surface error, the better, and the bias common to the whole area should also be small. Beam-level correction may be usable for applications that use many footprints together, such as training data for wall-to-wall canopy height maps \parencite{Potapov2021-yj,Lang2023-ic} and grid-cell statistics.

However, there are two broad limitations to extending these results to other regions and conditions. The first is that the areas where a DTM with a resolution of around 10 m exists are also limited. In tropical and subtropical forested countries, either there is no free nationwide DTM or, where a nationwide DTM exists, it is derived from synthetic aperture radar or photogrammetry and affected by the canopy, and in some countries it is available only for a fee. Furthermore, training a super-resolution model that can be transferred to the terrain of the target area requires 1 m-class training targets, and within the GEDI coverage, the countries where a free 1 m-class DTM covers most of the national territory are limited to some countries in Europe, the contiguous United States, New Zealand, and Japan. One way to solve this problem is to develop a 1 m-class super-resolution model that is transferable across the world and accepts an arbitrary input resolution at inference time. Super-resolution that outputs elevation at an arbitrary upscaling factor has been demonstrated by means of implicit neural representation \parencite{He2022-of,Yao2024-ry}, but both studies were validated over small areas, and we found no study that trained and validated 1 m-class super-resolution on a global scale. The second is that many factors can affect the results, and this study has shown only a few cases. The results of this study were obtained for forested areas that are steep by global standards, and the accuracy required of super-resolution and the results may change depending on the error distributions of the super-resolution and of the GEDI observed ground elevation. In particular, the required accuracy becomes more stringent for beams with few footprints and for footprints with large errors in the GEDI ground elevation \parencite{Li2025-oz}. The influence is considered to be greater in areas with small relief; in areas where matching by elevation differences is hard to determine uniquely even with a 1 m-resolution DTM, matching would be very difficult with a super-resolved surface \parencite{Tang2023-kw,Schleich2023-xz}. In addition, DTMs that differ from the GSI DTMs and FABDEM used as the inputs and the training target in production method, data representation, or vertical datum may give different results.

\section{Conclusion}

This study examined, in steep forested areas of western Japan, whether a 1 m-resolution reference DTM generated by super-resolution from a 10 m-resolution DTM and an auxiliary DTM can substitute for a high-resolution DTM (DEM1A) in the geolocation correction of GEDI footprints. In the super-resolution, training with only the 10 m-resolution DTM as input could not compensate for the elevation biases that vary from place to place, and mid-fusion of FABDEM, whose origin differs from that of the primary input, substantially improved the accuracy. In the beam-level correction, the super-resolved reference DTM gave almost the same offsets as DEM1A, and the degree of agreement increased with the number of footprints in a beam. In the footprint-level correction, the super-resolved reference DTM did not reproduce the DEM1A-based corrected positions.

We consider that this agreement was obtained in the beam-level correction because matching many footprints at once lowered the sensitivity of the offset to the reference-surface error. Beams that did not agree with DEM1A were more frequent among beams with few footprints. In contrast, because the footprint-level correction tries to resolve the elevation difference with a single footprint alone, the closer the reference-surface error is to the elevation difference to be resolved, the more the corrected position is dominated by the error. The reference DTM obtained in this study by super-resolving real data is not sufficiently accurate, and footprint-level correction using it is not a substitute for footprint-level correction with DEM1A.

Most of the error of the super-resolution model in this study lies at scales comparable to those that a 10 m-resolution DTM can represent, and remains even after averaging with the Gaussian window. Therefore, what is needed to obtain a super-resolution model that improves the correction accuracy is to reduce the error at these scales rather than to reproduce fine relief. The addition of information sources of different origins and the use of optical images as an information source that explains the error of the input DTM are tasks for future work. In addition, the super-resolution accuracy obtained with synthetic data is not a substitute for the performance on real data, and super-resolution used for geolocation correction must be evaluated on real data.

From the above, under similar conditions, a reference DTM obtained by fusing a DTM of roughly 10 m resolution with a global DTM and super-resolving it to 1 m resolution may be usable in place of a 1 m-resolution DTM in the beam-level correction. As long as correction with a 1 m-resolution DTM is effective, this could extend the areas where geolocation correction can be used, in applications that use many footprints together. However, the areas where both a DTM of roughly 10 m resolution as the super-resolution input and a 1 m-class DTM as the training target are available are limited, and the conditions under which the corrections agree have not been quantified either. To extend this method to areas without a 1 m-class DTM, the development of a super-resolution model that accepts an arbitrary input resolution and can be transferred to various regions is required, in addition to reducing the error of the input DTM.

\section*{Funding}
This work was supported by JSPS KAKENHI Grant Numbers JP24K00244 and JP26K14728.

\section*{Acknowledgments}
The 1 m and 10 m digital elevation models (DEM1A and DEM10B; Fundamental Geospatial Data, \url{https://service.gsi.go.jp/kiban/app/}), the geoid model Geoid 2024: Japan and its surroundings (JPGEO2024), and the height correction parameters for the 2025 height revision were provided by the Geospatial Information Authority of Japan and were processed by the authors. FABDEM V1-2 \parencite{Neal2023-fd} was provided by the University of Bristol under the CC BY-NC-SA 4.0 license (\url{https://creativecommons.org/licenses/by-nc-sa/4.0/}). FABDEM is produced using Copernicus WorldDEM-30 \textcopyright{} DLR e.V. 2010--2014 and \textcopyright{} Airbus Defence and Space GmbH 2014--2018 provided under COPERNICUS by the European Union and ESA; all rights reserved. The organisations in charge of the Copernicus programme by law or by delegation do not incur any liability for any use of the Copernicus WorldDEM-30. GEDI L2A Version 2 data were obtained from the NASA Land Processes Distributed Active Archive Center (LP DAAC). The land use and land cover data used in this paper were provided by JAXA (High-Resolution Land Use and Land Cover Map of Japan, version 25.04). This paper uses the following works with modification: DCHM (Digital Canopy Height Model) 0.5 m, Hyogo Prefecture, and the Kochi Prefecture forest resource data (measured in 2018), both published through the G-spatial Information Center under CC BY 4.0.

\section*{Data and code availability}
The code for reading the DTMs and FABDEM, for super-resolution, and for the geolocation correction, the trained weights of the super-resolution models, and the GEDI footprint positions before and after the correction with their observed elevations will be made publicly available on GitHub upon publication of the peer-reviewed version of this paper. The input data (GEDI L2A, FABDEM, the GSI digital elevation models, the JAXA land use and land cover map, and the DCHMs) are available from the providers listed in the Acknowledgments.

\section*{Declaration of competing interest}
The authors declare that they have no known competing financial interests or personal relationships that could have appeared to influence the work reported in this paper.

\section*{Declaration of generative AI use}
During the preparation of this work, the authors used Claude (Anthropic) through Claude Code to draft and translate text, to write and debug analysis code, and to typeset the manuscript. The authors reviewed and edited all output and take full responsibility for the content of this paper.

\printbibliography[title={References}]

\clearpage
\appendix
\renewcommand{\thesection}{\Alph{section}}
\renewcommand{\thefigure}{\thesection.\arabic{figure}}
\renewcommand{\thetable}{\thesection.\arabic{table}}

\section{Model details}

This appendix details, for the super-resolution model described in Section 4.1, the processing that produces the inputs to the model and the configuration of each layer of the network. The input processing consists of expanding the base elevation onto the grids of the primary input and the output, and resampling the auxiliary DTM onto the grid of the primary input.

The expansion of the base elevation, the Global skip $U$, and each upsampling stage of the progressive upsampling all use bilinear interpolation with the same convention. That is, the center of each output pixel is mapped to coordinates on the input grid, the value is interpolated at that position, and outside the input grid the values of the edge pixels are extended. Since pixel $n$ of the 10 m grid of the primary input exactly covers pixels $10n$ to $10n+9$ of the 1 m grid of DEM1A (Appendix B), the center of a pixel of the 10 m grid corresponds to coordinate $10n+4.5$ on the 1 m grid, which is consistent with the pixel-center mapping of this interpolation.

The base elevation $z_b$ is the mean of the primary input in each of the 5$\times$5 cells obtained by dividing the patch every 10$\times$10 pixels of the primary input. $\tilde{z}_b$ is this value expanded onto the grid of the primary input (50$\times$50) by bilinear interpolation with the centers of the cells as sample points, and $\hat{z}_b=U(\tilde{z}_b)$ is $\tilde{z}_b$ expanded onto the output grid (500$\times$500) by the same interpolation $U$ as the Global skip. Since $U$ is linear, the prediction with the network output set to 0 is $k\,U(\bar{z}_{lr})+\hat{z}_b=U(z_{lr}-\tilde{z}_b)+U(\tilde{z}_b)=U(z_{lr})$, which coincides with the surface obtained by bilinearly interpolating the primary input as it is. Therefore, the network $f$ does not need to cancel out the component arising from the way the base elevation is expanded, and it learns only the amount to be added to the bilinear interpolation of the primary input. When the primary input is DEM10B, as in (M1) and (M2), this surface is (M0) in the Results section, and when the primary input is synthetic data, as in (M3) and (M4), it is the bilinear-interpolation surface of the synthetic data in Table 5.2.

FABDEM, the auxiliary DTM, is resampled onto the grid of the primary input before being fed to the auxiliary-DTM encoder. FABDEM is a grid with 1 arcsecond spacing whose sample points lie on integer arcseconds, and since the edges and the center of a patch are placed on integer arcseconds (Appendix B), 21$\times$21 samples, including those on the edges, fall within one patch. The center of pixel $n$ ($n=0,\ldots,49$) of the primary-input grid is located at position $0.4n+0.2$ in terms of the index of the FABDEM samples. The value of each pixel is obtained by bilinear interpolation from the neighboring 2$\times$2 samples, but missing samples are excluded from the weights, and the result is normalized by dividing by the sum of the remaining weights. This sum of the weights of the valid samples (0 to 1) is used as the confidence channel $T$. Pixels with $T=0$, that is, pixels whose neighboring samples are all missing, are filled with $\tilde{z}_b$ (the deviation becomes 0). If all samples are valid, this interpolation coincides with ordinary bilinear interpolation.

The configuration of each layer of the network is shown in Table A.1. All convolutions have stride 1, and the 3$\times$3 convolutions use reflect padding to output feature maps of the same size as their input. The negative slope of LeakyReLU is 0.2. The configurations of (M2) and (M4), which do not use the auxiliary DTM, have neither the auxiliary-DTM encoder nor the fusion layer. The number of parameters of (M1), which uses the auxiliary DTM, is 1.9M in total.

\begin{table*}[t]
\centering
\small
\caption{Layer configuration of the super-resolution model. The output size is given as (height, width, number of channels), for an input patch of 50$\times$50 pixels on the primary-input grid. The numbers in parentheses in the Layers column are the numbers of input and output channels.}
\begin{adjustbox}{max width=\textwidth}
\begin{tabular}{p{0.197\linewidth}p{0.444\linewidth}p{0.279\linewidth}}
\toprule
Component & Layers & Output size \\
\midrule
Primary-input encoder $E_{\mathrm{main}}$ & 3$\times$3 convolution (1$\to$32) + ReLU, 3$\times$3 convolution (32$\to$64) + ReLU, 3$\times$3 convolution (64$\to$64) + ReLU & (50, 50, 64) \\
Auxiliary-DTM encoder $E_{\mathrm{aux}}$ & 3$\times$3 convolution (2$\to$32) + ReLU, 3$\times$3 convolution (32$\to$64) + ReLU, 3$\times$3 convolution (64$\to$64) + LeakyReLU & (50, 50, 64) \\
Fusion $G$ & channel-wise concatenation of the two feature maps (64+64), 3$\times$3 convolution (128$\to$64) + ReLU & (50, 50, 64) \\
RCAB & 3$\times$3 convolution (64$\to$64) + ReLU, 3$\times$3 convolution (64$\to$64), channel attention, sum with the input of the RCAB & (50, 50, 64) \\
Channel attention (part of RCAB) & global average pooling, 1$\times$1 convolution (64$\to$4) + ReLU, 1$\times$1 convolution (4$\to$64) + Sigmoid, channel-wise product of the obtained weights and the feature map & (50, 50, 64) \\
Residual group & 10 RCABs, 3$\times$3 convolution (64$\to$64), sum with the input of the residual group & (50, 50, 64) \\
RCAB body $B$ & 2 residual groups, 3$\times$3 convolution (64$\to$64), sum with the input of $B$ (long skip) & (50, 50, 64) \\
Progressive upsampling $P$ & in each of the three upsampling stages, bilinear interpolation (2.5$\times$, 2$\times$, 2$\times$) and 3$\times$3 convolution (64$\to$64) + LeakyReLU & (125, 125, 64), (250, 250, 64), (500, 500, 64) \\
Output layer & 3$\times$3 convolution (64$\to$1) & (500, 500, 1) \\
Global skip $U$ & bilinear interpolation of the normalized primary input (10$\times$) & (500, 500, 1) \\
\bottomrule
\end{tabular}
\end{adjustbox}
\end{table*}

\section{Details of patch placement and training settings}

This appendix details the dataset and training described in Section 5.1: how the patch positions are determined, how the reference-DTM values are obtained in the test blocks, how the synthetic data are produced, and the details of the training settings.

The patch positions are determined so that the grids of the three DTMs are exactly aligned with one another. The grid spacings of DEM1A, DEM10B, and FABDEM are 0.04 arcseconds, 0.4 arcseconds, and 1 arcsecond, respectively (Section 3.2). The implementation uses integer coordinates in which one pixel of DEM1A (0.04 arcseconds) is one unit; one pixel of DEM10B corresponds to 10 units, and 1 arcsecond to 25 units. A patch is 500$\times$500 pixels of DEM1A (20 arcseconds square), and the patch edges are placed on DEM10B pixel boundaries that are also integer arcseconds, that is, at positions that are multiples of 50 units. In this way, 50$\times$50 pixels of DEM10B exactly cover the patch (pixel $n$ of DEM10B covers pixels $10n$ to $10n+9$ of DEM1A), and 21$\times$21 FABDEM samples fall on the patch, with samples lying on its edges and at its center. This correspondence was verified for all patches by integer arithmetic.

The training and validation patches are placed within the DEM1A tiles. Since DEM1A is provided as one tile per tertiary mesh (30 arcseconds $\times$ 45 arcseconds, 750$\times$1125 pixels), the tiles are first divided into three groups: candidates for training and validation, test tiles, and excluded tiles. Only tiles in which at least 90\% of the DEM1A pixels are valid and for which DEM10B and FABDEM are both available over the same extent are considered, and the group of each tile is decided by the coordinates of its center. Tiles whose center lies inside a test block are assigned to test, tiles whose center lies within the area obtained by extending the test block by 5 arcminutes in each of the latitude and longitude directions are excluded, and the remaining tiles are candidates for training and validation. The candidates are split by secondary mesh, and a randomly selected 15\% (125) of the 836 secondary meshes are assigned to validation. Within each tile, the patch centers are placed at intervals of 200 pixels so that the patches fit inside the tile, giving 2 rows $\times$ 4 columns, a total of 8 patches per tile. Adjacent patches overlap by 300 pixels. In tertiary meshes with an odd longitude index, the western edge lies at the center of a DEM10B pixel, so the patches are moved 25 pixels (1 arcsecond) to the east to satisfy the alignment above.

In the test blocks, on the other hand, the patches are placed on a grid that is not confined to tertiary meshes, so that the reference-DTM coverage required for geolocation correction is available without gaps over the entire block. The northwest corner of each patch is placed at a position that is a multiple of 400 units from the origin at latitude 90 degrees and longitude 0 degrees, and adjacent patches overlap by 100 pixels. Since the origins of the primary meshes also fall on this grid, the patches can be placed by the same rule in any block. The patches are arranged so that the areas more than 30 pixels inside the edges of the patches together cover the entire block, and therefore the outermost patches extend beyond the block. The criteria for accepting patches are the same as for the training and validation patches (Section 5.1.1).

Although adjacent patches in a test block overlap, the super-resolution output is used as it is, as the 500$\times$500 pixel surface of each patch, and the outputs of overlapping patches are not merged into a mosaic by averaging. In the geolocation correction, for each search, one patch is selected according to the origin of that search (the recorded position $x_0$ of each footprint at the beam level, and the position $x_1$ after the beam-level correction at the footprint level), and the output of the same patch is used for all candidates of that search. The patch selected is the one with the largest margin among the patches whose margin to their edges is at least the sum of the number of pixels corresponding to the search radius of 20 m and the Gaussian-window half-width of 15 pixels (around 34 pixels). The correction with DEM1A as the reference DTM also uses DEM1A over the same patch extent, by the same rule. The margin guaranteed by the patch placement is 30 pixels, but inside the blocks the patch with the largest margin has a margin of at least 50 pixels, and only the footprints near the perimeter of a block (approximately 0.18\%) cannot satisfy the required margin. These footprints are excluded from the correction. On the other hand, the DEM1A values used for the evaluation at the corrected positions are obtained directly from the DEM1A tiles, not from the patches.

The synthetic data used as the input of (M3) and (M4) are also produced for each of these patches. The synthetic data are a 50$\times$50 pixel surface produced by averaging the DEM1A of the patch over every 10$\times$10 pixels. Each pixel is valid when at least 90 of the corresponding 100 pixels are valid, and its value is the mean of the valid pixels. The base elevation $z_b$ is obtained by averaging only the valid pixels of the synthetic data in each cell. Cells with no valid pixel are filled with the mean of the base elevations of the other cells of the same patch. Invalid pixels of the synthetic data are filled with $\tilde{z}_b$, the expansion of the base elevation obtained in this way. The same FABDEM as in (M1) is used as the auxiliary DTM. Since the accuracy is evaluated on the valid pixels of DEM1A, the evaluated pixels are common to (M1) through (M4).

The details of the training settings described in Section 5.1.3 are as follows. The four models differ only in the input configuration, and all other settings are the same. The Adam coefficients are 0.9 and 0.999, and no weight decay is used. The last incomplete batch of each epoch is not used. The same data-augmentation transformation is applied to all inputs and the training target after normalization. After each epoch, the RMSE in meters was computed over all valid pixels of the validation patches, and the model from the epoch with the smallest RMSE was used. The selected epochs were epoch 60 for (M1) (validation RMSE 3.71 m), epoch 40 for (M2) (5.68 m), epoch 97 for (M3) (0.414 m), and epoch 95 for (M4) (0.416 m).

\section{Comparison with DTM fusion without training}

This appendix compares (M0) and (M1) with reference DTMs produced by fusing DEM10B and FABDEM without training a neural network. The purpose is to examine whether the accuracy of (M1) can be obtained merely by combining the two DTMs. The case in which FABDEM alone is used as the reference DTM is also compared (Section 3.2.2).

The reference DTMs compared are the three in Table C.1. Unlike (M0), all three are produced by subtracting from each DTM the bias it has relative to DEM1A. All of them are built on the grid of the primary input and enlarged to 1 m resolution by the same bilinear interpolation as (M0). $z_{lr}$ is DEM10B, $z_{aux}$ is FABDEM resampled onto the grid of the primary input in the same way as for the super-resolution input (Appendix A), and $b_{lr}$ and $b_{aux}$ are the biases of DEM10B and FABDEM relative to DEM1A over all the training patches. (F3) is the weighted average, which is the most widely used method in DEM fusion \parencite{Okolie2022-hs}. The weight $w$ is set to the value that minimizes the variance of the error of (F3) over the training patches. The obtained values are 1.17 m for $b_{lr}$, 2.85 m for $b_{aux}$, and 0.63 for $w$.

\begin{table}[t]
\centering
\small
\caption{Reference DTMs compared. The expressions give the values on the grid of the primary input.}
\begin{adjustbox}{max width=\columnwidth}
\begin{tabular}{lp{0.204\linewidth}p{0.686\linewidth}}
\toprule
Symbol & Reference DTM & Expression \\
\midrule
F1 & DEM10B & $z_{lr}-b_{lr}$ \\
F2 & FABDEM & $z_{aux}-b_{aux}$ \\
F3 & Weighted average of (F1) and (F2) & $w\,(z_{lr}-b_{lr})+(1-w)\,(z_{aux}-b_{aux})$ \\
\bottomrule
\end{tabular}
\end{adjustbox}
\end{table}
Table C.2 shows the results of the evaluation with the same test patches and metrics as in Table 5.2. The RMSE of (F1) is 5.51 m, and the decrease from (M0) is 0.19 m. The RMSE of (M2), whose only input is DEM10B, is 5.32 m. (F3) has an RMSE of 4.96 m and an SD Bias of 2.04 m, falling short of the 3.81 m and 1.64 m of (M1), which takes the same two DTMs as input. (F3) and (M1) differ in the method of fusing the two DTMs into a surface on the 1 m grid, and the RMSE difference of 1.15 m results from using the trained super-resolution model instead of fusion with a fixed weight and bilinear interpolation. This difference is the effect of the model as a whole, including location-dependent correction and nonlinear fusion, and is not solely the contribution of generating detail at 1 m.

Next, the same geolocation correction as in the main text was performed with the reference DTMs in Table C.1, and the beam-level offsets were compared with those obtained with DEM1A (Table C.3). The set of beams compared is constructed in the same way as in Table 5.6. The proportion of beams whose offset difference is within 5 m is 58.7\% for (F1), 62.9\% for (F2), and 71.4\% for (F3), all higher than the 56.2\% of (M0), but none of them reaches the 84.6\% of (M1). However, the results in Table C.3 were obtained, for each reference DTM, from the beams whose beam-level solution did not reach the rim with either that reference DTM or DEM1A, so the sets of beams compared for (F3) and (M1) are not the same.

\begin{table*}[t]
\centering
\small
\caption{Accuracy of the reference DTMs in Table C.1. The test patches and metrics are the same as in Table 5.2. The rows for (M0), (M2), and (M1) are taken from Table 5.2.}
\begin{adjustbox}{max width=\textwidth}
\begin{tabular}{lrrrrrr}
\toprule
Reference DTM & RMSE (m) & MAE (m) & Avg. Bias (m) & SD Bias (m) & RMSE-Slope ($^{\circ}$) & RMSE-Aspect ($^{\circ}$) \\
\midrule
M0 & 5.70 & 4.22 & 1.476 & 2.668 & 11.4 & 45.8 \\
F1 & 5.51 & 4.07 & 0.305 & 2.668 & 11.4 & 45.8 \\
F2 & 7.23 & 5.67 & 0.483 & 2.949 & 13.9 & 50.1 \\
F3 & 4.96 & 3.79 & 0.370 & 2.041 & 11.7 & 44.8 \\
M2 & 5.32 & 3.93 & -0.008 & 2.650 & 10.2 & 44.1 \\
M1 & 3.81 & 2.82 & 0.162 & 1.642 & 9.6 & 41.4 \\
\bottomrule
\end{tabular}
\end{adjustbox}
\end{table*}
\begin{table*}[t]
\centering
\small
\caption{Difference between the beam-level offsets obtained with the reference DTMs in Table C.1 and the offsets obtained in the correction with DEM1A. The proportions of beams whose difference is within 2.5 m, within 5 m, and at least 10 m, and the median difference are shown. The rows for (M0), (M2), and (M1) are taken from Table 5.6.}
\begin{adjustbox}{max width=\textwidth}
\begin{tabular}{lrrrrr}
\toprule
Reference DTM & Number of beams & Within 2.5 m (\%) & Within 5 m (\%) & At least 10 m (\%) & Median (m) \\
\midrule
M0 & 1,697 & 23.8 & 56.2 & 8.1 & 4.2 \\
F1 & 1,687 & 24.4 & 58.7 & 7.4 & 4.2 \\
F2 & 1,648 & 28.0 & 62.9 & 8.5 & 3.9 \\
F3 & 1,716 & 34.5 & 71.4 & 5.3 & 3.3 \\
M2 & 1,679 & 26.1 & 60.6 & 7.1 & 4.0 \\
M1 & 1,736 & 56.4 & 84.6 & 2.8 & 2.4 \\
\bottomrule
\end{tabular}
\end{adjustbox}
\end{table*}

\section{Results of using flow accumulation at the beam level}

In the main text, the beam-level correction was performed by grid search, and grid search and flow accumulation were compared for the footprint-level correction. At the footprint level, flow accumulation was better in terms of both canopy height and agreement with the DEM1A-based corrected positions. This appendix restricts the reference surfaces of interest to (M1) and DEM1A and compares the results obtained when flow accumulation is also used for the beam-level correction with those in the Results section.

The beam-level flow accumulation uses the mean cost $\bar{c}(\xi)$ of the footprints belonging to the beam as the cost map and selects the corrected position by the same procedure as the footprint-level flow accumulation. That is, the FD8 algorithm is applied to $\bar{c}(\xi)$, and the flow-weighted centroid of the cells whose flow accumulation is in the top 1\% is taken as the offset $u$ of the beam. Whether the solution reached the rim is judged by whether the candidate closest to the centroid lies on the rim of the candidate set. The candidate set of the search, the subsequent footprint-level correction (grid search and flow accumulation), and the classification of correction failures are the same as in the main text. For the same reference DTM, the comparison is made between the correction with grid search at the beam level and the correction with flow accumulation at the beam level. For each reference DTM, the sets of evaluated footprints (Section 4.2.4) of the beam-level correction, the footprint-level grid search, and the footprint-level flow accumulation in the correction with grid search at the beam level are denoted by $\mathcal{E}^{\mathrm{gs}}$, $\mathcal{E}^{\mathrm{gs}}_{\mathrm{gs}}$, and $\mathcal{E}^{\mathrm{gs}}_{\mathrm{fa}}$, respectively, and those of the correction with flow accumulation at the beam level are denoted in the same way with fa as the superscript. In the comparison of the corrected positions (Table D.1), as in Section 5.4.1, the evaluated footprints used for the comparison are determined for each correction stage. The beam-level offsets are compared for the beams whose solution did not reach the rim of the search disc with either of the two methods, and the footprint-level corrected positions are compared for the footprints in $\mathcal{E}^{\mathrm{gs}}_{\mathrm{gs}}\cap\mathcal{E}^{\mathrm{fa}}_{\mathrm{gs}}$ for grid search and in $\mathcal{E}^{\mathrm{gs}}_{\mathrm{fa}}\cap\mathcal{E}^{\mathrm{fa}}_{\mathrm{fa}}$ for flow accumulation. In the comparison of accuracy (Table D.2), the intersection of the six sets of evaluated footprints \[\adjustbox{max width=\linewidth}{$\displaystyle (\mathcal{E}^{\mathrm{gs}}\cap\mathcal{E}^{\mathrm{gs}}_{\mathrm{gs}}\cap\mathcal{E}^{\mathrm{gs}}_{\mathrm{fa}})\cap(\mathcal{E}^{\mathrm{fa}}\cap\mathcal{E}^{\mathrm{fa}}_{\mathrm{gs}}\cap\mathcal{E}^{\mathrm{fa}}_{\mathrm{fa}})$}\] is used as the set of evaluated footprints for all rows, including the no-correction rows.

First, in terms of the corrected positions, the beam-level offset $u$ was almost the same for the two methods (Table D.1). For DEM1A, of the 1,825 beams whose beam-level solution did not reach the rim with either method, the difference was within 2.5 m for 93.8\%, and the median difference was 0.9 m. For (M1) as well, 92.7\% of the 1,778 beams were within 2.5 m and the median difference was 1.0 m; the differences are mostly limited to the size of one pixel of the reference DTM. For the subsequent footprint-level corrected positions, with flow accumulation, 79.4\% for DEM1A and 79.2\% for (M1) were within 2.5 m and the median distance was 1.2 m for both, which was comparable to the difference in the beam-level offsets. With grid search, the median distance was 9.0 m for DEM1A and 8.8 m for (M1). In grid search, the candidates with elevations close to the observed ground elevation lie along the contour lines of the reference surface (Section 5.4.1), so a movement of the origin of only about 1 m is enough for a distant candidate to be selected.

Even when the beam-level correction is changed to flow accumulation, the accuracy of elevation and canopy height is almost the same as with grid search. On the 208,706 footprints for DEM1A and the 191,349 footprints for (M1) that were evaluated in Table D.2, the differences between the two methods in the MAE of elevation and canopy height were at most 0.01 m for elevation and at most 0.005 m for canopy height at every correction stage. The only difference between the two methods is in the proportion of beams whose beam-level solution reached the rim of the search disc: with grid search it was 14.2\% (303 of 2,129 beams) for DEM1A and 16.4\% (350 beams) for (M1), whereas with flow accumulation it was 7.0\% (150 beams) and 8.5\% (182 beams), respectively.

In summary, changing the beam-level method from grid search to flow accumulation reduces the number of beams whose solution reaches the rim, but the corrected positions of the beams whose solution did not reach the rim in either correction remain almost unchanged, and the accuracy of elevation and canopy height of the footprints also remains almost unchanged. Therefore, in the main text, the same grid search as in the group-level correction of \textcite{Li2025-oz} was used at the beam level.

\begin{table*}[t]
\centering
\small
\setlength{\tabcolsep}{3.5pt}
\caption{Comparison of the corrected positions between the two methods at the beam level. For the same reference DTM, the distance between the corrected position with flow accumulation at the beam level and that with grid search (at the beam level, the difference in the per-beam offset $u$; at the footprint level, the distance between the corrected positions $x_2$) is summarized by the proportions within 0.5 m, within 2.5 m, within 5 m, and at least 10 m, the median, and the number of units (N, beams or footprints).}
\begin{adjustbox}{max width=\textwidth}
\begin{tabular}{llrrrrrr}
\toprule
Reference DTM & Stage & N & Within 0.5 m (\%) & Within 2.5 m (\%) & Within 5 m (\%) & At least 10 m (\%) & Median (m) \\
\midrule
DEM1A & beam-level & 1,825 & 21.6 & 93.8 & 98.4 & 0.4 & 0.9 \\
 & footprint-level (grid search) & 209,308 & 4.1 & 17.1 & 31.0 & 46.0 & 9.0 \\
 & footprint-level (flow accumulation) & 279,188 & 20.9 & 79.4 & 96.2 & 0.3 & 1.2 \\
\midrule
M1 & beam-level & 1,778 & 18.2 & 92.7 & 98.5 & 0.2 & 1.0 \\
 & footprint-level (grid search) & 192,094 & 4.3 & 17.6 & 31.6 & 45.2 & 8.8 \\
 & footprint-level (flow accumulation) & 264,808 & 21.7 & 79.2 & 95.8 & 0.4 & 1.2 \\
\bottomrule
\end{tabular}
\end{adjustbox}
\end{table*}
\begin{table}[t]
\centering
\small
\caption{Comparison of accuracy between the beam-level methods. For the same reference DTM, the MAE of elevation and canopy height of the two corrections, with grid search and with flow accumulation at the beam level, is shown in the form ``grid search / flow accumulation''. The canopy-height MAE was computed in the five blocks where the DCHM is available, for the footprints that have a DCHM value of at least 2 m within a radius of 12.5 m (approximately 60,000 for DEM1A and approximately 56,000 for (M1)).}
\begin{adjustbox}{max width=\columnwidth}
\begin{tabular}{llll}
\toprule
Reference DTM & Correction stage & Elevation MAE (m) & Canopy-height MAE (m) \\
\midrule
DEM1A & No correction & 4.60 / 4.60 & 5.26 / 5.26 \\
 & beam-level & 3.60 / 3.61 & 5.16 / 5.16 \\
 & beam-level+ footprint-level(gs) & 0.05 / 0.05 & 4.98 / 4.98 \\
 & beam-level+ footprint-level(fa) & 0.92 / 0.92 & 4.93 / 4.93 \\
\midrule
M1 & No correction & 4.77 / 4.77 & 5.38 / 5.38 \\
 & beam-level & 3.83 / 3.83 & 5.29 / 5.29 \\
 & beam-level+ footprint-level(gs) & 2.57 / 2.58 & 5.21 / 5.21 \\
 & beam-level+ footprint-level(fa) & 2.59 / 2.59 & 5.15 / 5.15 \\
\bottomrule
\end{tabular}
\end{adjustbox}
\end{table}

\section{Agreement of canopy height and ground elevation when footprints are shifted}

This appendix presents the results of examining the response of canopy height and ground elevation to position, discussed in Section 6.3, using the correction records with DEM1A as the reference DTM. The analysis covers the 65,856 footprints for which the ground elevation and canopy height were obtained at all 33 positions described below, out of the 66,390 footprints that could be evaluated in all four types of correction results in the five test blocks where the DCHM is available.

First, in the same way as \textcite{Quiros2021-mu} and \textcite{Kutchartt2022-lz}, all footprints are shifted uniformly from their recorded positions. The shifts are in eight directions at 45-degree intervals clockwise from the along-track direction (0 degrees), at distances of 5 m, 10 m, 15 m, and 20 m, which gives 33 positions including the recorded position. At each position, the ground-elevation residual, obtained by subtracting the Gaussian-weighted elevation of DEM1A from the observed ground elevation of GEDI, and the canopy-height error, obtained by subtracting the maximum of the DCHM within a radius of 12.5 m from RH98, are computed. Table E.1 shows the MAE for each distance. The values at a distance of 0 m in Tables E.1 and E.2 differ from the no-correction values in Tables 5.3 and 5.4 because the sets of footprints differ.

\begin{table}[t]
\centering
\small
\caption{MAE of the canopy-height error and the ground-elevation residual when the footprints are shifted uniformly from their recorded positions. For each distance, the MAE over all footprints was computed for each of the eight directions, and the minimum and maximum are shown. A distance of 0 m is the recorded position.}
\begin{adjustbox}{max width=\columnwidth}
\begin{tabular}{rll}
\toprule
Distance (m) & Canopy-height MAE (m) & Ground-elevation MAE (m) \\
\midrule
0 & 5.34 & 5.17 \\
5 & 5.36--5.40 & 5.34--5.64 \\
10 & 5.45--5.52 & 6.11--6.67 \\
15 & 5.58--5.67 & 7.34--8.08 \\
20 & 5.74--5.85 & 8.87--9.69 \\
\bottomrule
\end{tabular}
\end{adjustbox}
\end{table}
Next, for each footprint, the position with the smallest absolute canopy-height error (``selected by canopy height'' in Table E.2) or the position with the smallest absolute ground-elevation residual (``selected by ground elevation'') was chosen from the same 33 positions, and the MAE of the canopy-height error and the ground-elevation residual at the chosen positions was computed (Table E.2). With maximum distances of 5 m, 10 m, 15 m, and 20 m, the candidates are 9, 17, 25, and 33 positions, respectively. When positions are selected by the canopy-height error, the position is chosen by the very quantity used for the evaluation, so the canopy-height error decreases as the number of candidates increases. Even when the MAEs were computed separately for each of the five test blocks, with a maximum distance of at least 15 m, selecting by canopy height made the ground-elevation MAE larger than at the recorded position, and selecting by ground elevation reduced the canopy-height MAE relative to the recorded position by less than 0.5 m.

\begin{table}[t]
\centering
\small
\setlength{\tabcolsep}{3.5pt}
\caption{MAE of the canopy-height error and the ground-elevation residual when a position is selected for each footprint. A maximum distance of 0 m is the recorded position.}
\begin{adjustbox}{max width=\columnwidth}
\begin{tabular}{rrrrr}
\toprule
 & \multicolumn{2}{c}{Selected by canopy height} & \multicolumn{2}{c}{Selected by ground elevation} \\
\cmidrule(lr){2-3} \cmidrule(lr){4-5}
\shortstack{Maximum\\distance (m)} & \shortstack{Canopy-height\\MAE (m)} & \shortstack{Ground-elevation\\MAE (m)} & \shortstack{Canopy-height\\MAE (m)} & \shortstack{Ground-elevation\\MAE (m)} \\
\midrule
0 & 5.34 & 5.17 & 5.34 & 5.17 \\
5 & 4.22 & 5.00 & 5.15 & 2.64 \\
10 & 3.45 & 5.15 & 5.06 & 1.27 \\
15 & 2.91 & 5.60 & 5.03 & 0.64 \\
20 & 2.54 & 6.32 & 5.05 & 0.40 \\
\bottomrule
\end{tabular}
\end{adjustbox}
\end{table}

\end{document}